\documentclass{iopjournal}
\usepackage{amsmath}           		          	
\usepackage{graphicx,epstopdf}					
\usepackage{amssymb}
\pdfoutput = 1 
\allowdisplaybreaks

\begin{document}
\articletype{paper}
\title{A simple model of current ramp down in the ITER tokamak}
\author{Richard Fitzpatrick\orcid{0000-0001-6237-9309}}
\affil{Institute for Fusion Studies, Department of Physics, University of Texas at Austin, Austin TX 78712, USA}
\email{rfitzp@utexas.edu}

\begin{abstract}
The controlled  ramp down of the toroidal plasma current in the ITER tokamak is simulated using a simple model that employs cylindrical geometry. The magnetohydrodynamical  (MHD) stability of the plasma throughout the whole current ramp is also  calculated. 
The only potentially unstable MHD mode is  the $m=2/n=1$ classical  tearing mode. The envisioned 60 second ramp down of the plasma current in ITER is
found to be perfectly feasible, provided that the plasma is sufficiently hot at the start of the ramp. However, attempts to ramp down the current on a significantly 
faster time scale are predicted to excite 2/1 tearing modes that are likely to lock to the vacuum vessel, and trigger a disruption. 

\end{abstract}

\section{Introduction}\label{s1}
In essence, a tokamak is a   transformer in which a time-varying current flowing through a central solenoid induces a toroidal current in 
the surrounding plasma, which constitutes a single-turn secondary winding \cite{book}. Tokamaks differ from conventional transformers principally  in terms of
their sheer scale. For example, the poloidal magnetic energy contained in the ITER tokamak, when carrying its designed toroidal current of 15 MA, is approximately
2 GJ. At the end of  the discharge, this energy must be safely disposed of by means of a controlled ramp down of the plasma current \cite{jack,imb,vries,asdex,demo,tcv,muld}. 
The major obstacles to the safe shut-down of the plasma discharge are macroscopic magnetohydrodynamical (MHD) instabilities that can be excited as the
toroidal current and safety-factor profiles evolve during the ramp down \cite{fried}, and which may trigger a major disruption \cite{kad,sweeney,jet}. The most dangerous instabilities are ideal external-kink modes
and tearing modes. The main aim of this paper is to simulate a current ramp down in ITER, while simultaneously determining whether or not the discharge is MHD stable. 
For the sake of simplicity, and speed of calculation, the analysis is performed in cylindrical geometry. Indeed, the analysis is a straightforward extension of
that presented in an earlier paper \cite{rf1}.

 Previous calculations have attempted  to assess the MHD stability of a cylindrical plasma solely in terms of the location
of the discharge in $q_a$-$l_i$  space, where $q_a$  is the safety-factor at the plasma boundary, and $l_i$  the normalized plasma internal self-inductance \cite{cheng}.
However, such calculations make use of highly contrived model current profiles that do not necessarily bear any relation to those found in actual tokamaks. 
In fact, conventional $q_a$-$l_i$  stability diagrams are known to provide only a heuristic characterization of tokamak stability. In particular, MHD stability depends sensitively on the detailed current and pressure profiles, which cannot be uniquely parameterized by $q_a$ and $l_i$ alone \cite{fried, hus}.
Hence, in this paper, we intend to perform a more self-consistent assessment of plasma stability during current ramp downs. 

\section{Fundamental model}
\subsection{Coordinates}
Let us model our tokamak plasma, in the simplest possible way imaginable,  as a periodic cylinder of circular cross-section. Let $r$, $\theta$, $z$ be conventional cylindrical coordinates, and
let the magnetic axis of the plasma correspond to $r=0$. The system is assumed to be periodic in $z$ with period length $2\pi\,R_0$, where $R_0$ is the
simulated major radius of the plasma. Let $a$ be the (initial) minor radius of the plasma, and $\epsilon=a/R_0$ the (initial) inverse aspect-ratio.

\subsection{Electromagnetic equations}
The equilibrium magnetic field is represented as
\begin{equation}
{\bf B}(r,t) = B_\theta(r,t)\,{\bf e}_\theta+ B_0\,{\bf e}_z,
\end{equation}
where $B_0$ is the constant toroidal magnetic field generated by  steady currents flowing in external  magnetic field-coils. The poloidal field, $B_\theta(r,t)$, on the other hand, is generated
via transformer action by the time variation of the current flowing in the tokamak's central solenoid. The current density in the plasma is written ${\bf j}= j_z\,{\bf e}_z$. Moreover, the Amp\`{e}re-Maxwell equation (neglecting the displacement current) implies that
\begin{equation}\label{e2}
\mu_0\,j_z(r,t) = \frac{1}{r}\,\frac{\partial}{\partial r}(r\,B_\theta).
\end{equation}
The total toroidal current carried by the plasma is
\begin{equation}\label{ei}
I_p(t)= \int_0^a 2\pi\,r\,j_z\,dr = \frac{2\pi\,a\,B_\theta(a,t)}{\mu_0},
\end{equation}
The plasma safety-factor profile takes the form 
\begin{equation}
q(r,t) = \frac{r\,B_0}{R_0\,B_\theta},
\end{equation}
while the electric field in the plasma is written ${\bf E}= E_z\,{\bf e}_z$. The Faraday-Maxwell equation yields
\begin{equation}\label{e5}
\frac{\partial E_z}{\partial r} = \frac{\partial B_\theta}{\partial t},
\end{equation}
whereas Ohm's law gives
\begin{equation}\label{e6}
E_z = \eta\,j_z
\end{equation}
where
\begin{equation}
\eta = \frac{Z\,\ln{\mit\Lambda}}{1.96\,6\sqrt{2}\,\pi^{3/2}}\,\frac{m_e^{\,1/2}\,e^{\,2}\,c^{\,4}\,\mu_0^{\,2}}{T_e^{\,3/2}}
\end{equation}
is the Spitzer plasma resistivity \cite{spitzer,fitz}. Here, $T_e(r,t)$ is the electron temperature profile, $Z$ the effective ion charge number, $\ln{\mit\Lambda}\simeq 15$ the
Coulomb logarithm, $m_e$ the electron mass, $e$ the magnitude of the electron charge, $c$ the velocity of light in vacuum, and
$\mu_0$ the vacuum permeability. 
Note that, for the sake of simplicity, we are neglecting both the neoclassical enhancement of the plasma resistivity, as well as 
the non-inductive bootstrap current \cite{fitz1}. The neglect of the bootstrap current  is fairly reasonable in the late ramp-down phase  of a tokamak plasma, during which temperature and pressure gradients are greatly reduced from their peak values. The neoclassical  enhancement could most easily be taken into account in our analysis by multiplying $Z$
 by the volume-averaged enhancement factor. 
  
\subsection{Thermal transport}
 The electron thermal transport equation takes the form\,\cite{fitz}
\begin{equation}\label{e7}
\frac{3}{2}\,n_e\,\frac{\partial T_e}{\partial t} = n_e\,\frac{1}{r}\,\frac{\partial}{\partial r}\left(r\,\chi_\perp\,\frac{\partial T_e}{\partial r}\right)
 +\eta\,j_z^{\,2} + p_{\rm aux},
 \end{equation}
 where $n_e$ is the electron number density,  $\chi_\perp(r)$ is the electron perpendicular energy diffusivity due to small-scale plasma turbulence, and
 $p_{\rm aux}(r,t)$ the rate of auxiliary heating per unit volume. 
 Note that, for the sake of simplicity, we are treating $n_e$ as a spatial and temporal constant. Moreover, we expect $\chi_\perp \sim 1\,{\rm m^2\,s}^{-1}$ \cite{book}.
 Incidentally, the estimate $\chi_\perp \sim 1\,{\rm m^2\,s}^{-1}$ for ITER is reasonably consistent with the ITER98(y,2) scaling law \cite{iter}.
 
\subsection{Energy balance}
 The following poloidal magnetic energy balance equation can be derived from Eqs.~(\ref{e2}), (\ref{e5}), and (\ref{e6}) \cite{muld,ejima}:
 \begin{equation}\label{e8}
 \frac{dW_i}{dt} = {\mit\Gamma}_m - P_{\rm oh},
 \end{equation}
 where
 \begin{equation}
 W_i = 4\pi^2\,R_0\int_0^a\frac{B_\theta^{\,2}}{2\,\mu_0}\,r\,dr
 \end{equation}
 is the poloidal magnetic energy associated with the plasma current in the region internal to the plasma, 
 \begin{equation}
 {\mit\Gamma}_m=  \frac{4\pi^2\,R_0\,a}{\mu_0}\,(B_\theta\,E_z)_{r=a}
 \end{equation}
 the inward flux of poloidal magnetic energy across the plasma boundary, and
 \begin{equation}
 P_{\rm oh} = 4\pi^2\,R_0\int_0^a\eta\,j_z^{\,2}\,r\,dr
 \end{equation}
  the rate of poloidal magnetic energy dissipation due to ohmic heating within the plasma.
 
 We can also write
 \begin{equation}\label{e10}
 \frac{dW_e}{dt} = {\mit\Gamma}_s - {\mit\Gamma}_m,
 \end{equation}
 where
 \begin{equation}
 W_e = \frac{1}{2}\,L_e\,I_p^{\,2}
 \end{equation}
 is the poloidal magnetic energy associated with the plasma current in the region external to the plasma, 
 \begin{equation}
 L_e = \mu_0\,R_0\,l_e
 \end{equation}
 the external plasma self-inductance,
 \begin{equation}
 l_e=\ln\left(\frac{8}{\epsilon}\right)-2
 \end{equation}
 the normalized external self-inductance \cite{fried}, and ${\mit\Gamma}_s$ the inward flux of electromagnetic energy from the
 central solenoid.
 Equations~(\ref{e8}) and (\ref{e10}) yield 
\begin{equation}\label{erty}
{\mit\Gamma}_s = \frac{dW}{d\hat{t}} + P_{\rm oh},
\end{equation}
where $W=W_i+W_e$ is the total poloidal magnetic energy. 
Thus, the electromagnetic energy supplied  by the central solenoid is either stored in the plasma's poloidal magnetic field, or is
dissipated ohmically. 
 
 The following thermal energy balance equation can be
 derived from Eq.~(\ref{e7}):
 \begin{equation}\label{eax}
 \frac{dW_{\rm th}}{dt} = {\mit\Gamma}_{\rm th} + P_{\rm oh} + P_{\rm aux},
 \end{equation}
 where
 \begin{equation}
 W_{\rm th} = 4\pi^2\,R_0\int_0^a \frac{3}{2}\,n_e\,T_e\,r\,dr
 \end{equation}
 is the net thermal energy within the plasma, 
 \begin{equation}
 {\mit\Gamma}_{\rm th} = 4\pi^2\,R_0\,n_e\left(\chi_\perp\,r\,\frac{\partial T_e}{\partial r}\right)_{r=a}
 \end{equation}
 the inward flux of thermal energy across the plasma boundary,
 and
 \begin{equation}
 P_{\rm aux}(t)= 4\pi^2\,R_0\int_0^a p_{\rm aux}(r,t)\,r\,dr
 \end{equation}
 the net auxiliary heating rate. 
 
\subsection{Circuit equation}
We can write the following equivalent  circuit equation for the plasma \cite{romero}:
\begin{equation}\label{e20}
0= I_p\,{\cal R}_p + \frac{d(L_p\,I_p)}{dt} +\frac{d (M_{ps}\,I_s)}{dt}.
\end{equation}
 Here, ${\cal R}_p$ is the electrical resistance of the plasma, $L_p= L_i+L_e$ the total self-inductance of the plasma, $M_{ps}$ the mutual inductance between the
 plasma and the central solenoid, and $I_s$ the current flowing in the central solenoid. 
 Moreover,
 \begin{equation}
 L_i = \frac{W_i}{(1/2)\,I_p^{\,2}}= \frac{1}{2}\,\mu_0\,R_0\,l_i,
 \end{equation}
 where
 \begin{equation}
 l_i = \frac{2}{[a\,B_\theta(a)]^2}\int_0^a B_\theta^{\,2}(r)\,r\,dr
 \end{equation}
 is the normalized internal self-inductance of the plasma \cite{fried}. The loop-volts are defined as 
 \begin{equation}\label{circuit}
 V_{\rm loop} = I_p\,{\cal R}_p + \frac{d (L_p\,I_p)}{dt} = -\frac{d(M_{ps}\,I_s)}{dt}.
 \end{equation}
 
 Equation~(\ref{e20}) yields the poloidal magnetic energy conservation equation 
 \begin{equation}\label{e25c}
 \frac{dW}{dt}  = -P_{\rm oh} + {\mit\Gamma}_s,
 \end{equation}
 where  
 \begin{equation}
 {\cal R}_p = \frac{P_{\rm oh}}{I_p^{\,2}},
 \end{equation}
 and
 \begin{equation}
 {\mit\Gamma}_s= V_{\rm loop}\,I_p.
 \end{equation}
 Note that Eq.~(\ref{e25c}) is consistent with Eq.~(\ref{erty}). 
 
\subsection{Scale quantities}
 Let  
 \begin{equation}
 B_{\theta\,a}=\frac{\epsilon\,B_0}{q_a}
 \end{equation}
  be 
a typical poloidal magnetic field-strength. Here, $q_a$ is the edge safety-factor at the start of the current ramp.
Suppose that $\chi_\perp (r)= \chi_0\,\hat{\chi}(r)$, where $\chi_0$ is a
 typical value of the perpendicular energy diffusivity, and $\hat{\chi}(r)$ is dimensionless and of order unity. 
 We can define \cite{rf1}
 \begin{align}
 T_0\equiv \frac{\eta(T_0)\,B_{\theta\,a}^{\,2}}{\mu_0^{\,2}\,n_e\,\chi_0}=9.95\times 10^{-1}\,Z^{\,2/5}\,{\mit\Lambda}_{15}^{\,2/5}\,n_{20}^{-2/5}\,\chi_0^{-2/5}\,B_{\theta\,a}^{\,4/5}\,\,[{\rm keV}]
 \end{align}
 as the typical electron temperature. Here, ${\mit\Lambda}_{15}=\ln{\mit\Lambda}/15$ and $n_{20}=n_e/10^{\,20}$. 
 All other quantities are in SI units. Likewise, 
 \begin{equation}\label{taur}
 \tau_R \equiv \frac{a^2\,\mu_0}{\eta(T_0)}= 4.99\times 10^1\,a^2\,Z^{-2/5}\,{\mit\Lambda}_{15}^{-2/5}\,n_{20}^{-3/5}\,\chi_0^{-3/5}\,B_{\theta\,a}^{\,6/5}\,\,[{\rm s}]
 \end{equation}
 is the conventional resistive diffusion time, whereas 
  $\tau_c= a^2/\chi_0\,\,[{\rm s}]$
 is the energy confinement time. The central plasma poloidal beta is defined
 \begin{equation}\label{e16}
 \beta_p\equiv \frac{\mu_0\,n_e\,T_0}{B_{\theta\,a}^{\,2}} =\frac{\tau_c}{\tau_R}=  2.00\times 10^{-2}\,Z^{\,2/5}\,{\mit\Lambda}_{15}^{\,2/5}\,n_{20}^{3/5}\,\chi_0^{-2/5}\,B_{\theta\,a}^{-6/5}.
 \end{equation}
 The typical toroidal plasma current is
$I_0 =2\pi\,a\,B_{\theta\,a}/\mu_0= 5.00\,a\,B_{\theta\,a}\,\,[{\rm MA}]$,  whereas the typical inductive electric field-strength in the plasma takes the form
 \begin{equation}
 E_0  \equiv \frac{\beta_p\,\chi_0\,B_{\theta\,a}}{a}=\frac{a\,B_{\theta\,a}}{\tau_R}= 2.00\times 10^{-2}\,a^{-1}\,Z^{\,2/5}\,{\mit\Lambda}_{15}^{\,2/5}\,n_{20}^{3/5}\,\chi_0^{\,3/5}\,B_{\theta\,a}^{-1/5}\,\,[{\rm V\,m}^{-1}].
 \end{equation}
 Finally, the typical magnetic energy is $W_0=\mu_0\,R_0\,I_0^{\,2}$, the typical plasma resistance is ${\cal R}_0=\mu_0\,R_0/\tau_R$, and the typical loop-volts
 is $V_0=I_0\,{\cal R}_0$. 
 
\subsection{Normalization}
Let 
 \begin{align}
 t&=\tau_R\,\hat{t},\\[0.5ex]
 r&= a\,\delta(\hat{t})\,\hat{r},\\[0.5ex]
 B_\theta(r,t) &= \frac{B_{\theta\,a}\,\hat{B}(\hat{r},\hat{t})}{\delta(\hat{t})},
 \end{align}
 where $\delta(0)=1$. Here, $\delta(\hat{t})\,a$ is the plasma minor radius at normalized time $\hat{t}$. 
 It follows that
 \begin{align}
 \left.\frac{\partial}{\partial r}\right|_t&= \frac{1}{a\,\delta}\left.\frac{\partial}{\partial \hat{r}}\right|_{\hat{t}},\\[0.5ex]
 \left.\frac{\partial}{\partial t}\right|_r&= \frac{1}{\tau_R}\left(\left.\frac{\partial}{\partial \hat{t}}\right|_{\hat{r}}-V\,\hat{r}\left.\frac{\partial}{\partial \hat{r}}\right|_{\hat{t}}\right),
 \end{align}
 where
 \begin{equation}
 V(\hat{t}) = \frac{d\ln\delta}{d\hat{t}}.
 \end{equation}

 Let us adopt the following convenient normalization scheme:  $T_e(r,t)=T_0\,\hat{T}(\hat{r},\hat{t})$, $B_\theta(r,t)=B_{\theta\,a}\,\hat{B}(\hat{r},\hat{t})$,  $E(r,t)=E_0\,\hat{E}(\hat{r},\hat{t})$, $j_z(r,t)=[B_{\theta\,a}/(\mu_0\,a)]\,\skew{3}\hat{j}(\hat{r},\hat{t})$,  $q(r,t)=q_a\,\hat{q}(\hat{r},\hat{t})$, 
 $p_{\rm aux}(r,t)= [B_{\theta\,a}^{\,2}/(\mu_0\,\tau_R)]\,\hat{p}_{\rm aux}(\hat{r},\hat{t})$, 
 $I_p(t)=I_0\,\hat{I}_p(\hat{t})$,   $W_i(t)=W_0\,\hat{W}_i(\hat{t})$,  $W_e(t)=W_0\,\hat{W}_e(\hat{t})$, $W_{\rm th}(t)=W_0\,\hat{W}_{\rm th}(\hat{t})$, $P_{\rm oh}(t) = (W_0/\tau_R)\hat{P}_{\rm oh}(\hat{t})$, $P_{\rm aux}(t)= (W_0/\tau_R)\hat{P}_{\rm aux}(\hat{t})$, 
 ${\mit\Gamma}_m(t) = (W_0/\tau_R)\,\hat{\mit\Gamma}_m(\hat{t})$, ${\mit\Gamma}_s(t) = (W_0/\tau_R)\,\hat{\mit\Gamma}_s(\hat{t})$, ${\mit\Gamma}_{\rm th} (t)= (W_0/\tau_R)\,\hat{\mit\Gamma}_{\rm th}(\hat{t})$, ${\cal R}_p(t) ={\cal R}_0\,\hat{\cal R}_p(\hat{t})$, and $V_{\rm loop}(t) = V_0\,\hat{V}_{\rm loop}(\hat{t})$. 
 
It follows that 
 \begin{align}
\frac{\partial \hat{B}}{\partial\hat{t}} &= \frac{\partial\hat{E}}{\partial\hat{r}} + V\,\frac{\partial (\hat{r}\,\hat{B})}{\partial\hat{r}},\label{e85}\\[0.5ex]
 \skew{3}\hat{j}(\hat{r},\hat{t}) &= \frac{1}{\delta^2\,\hat{r}}\,\frac{\partial\,(\hat{r}\,\hat{B})}{\partial\hat{r}},\label{jdef}\\[0.5ex]
 \hat{E}(\hat{r},\hat{t}) &= \frac{\skew{3}\hat{j}(\hat{r},\hat{t})}{\hat{T}^{\,3/2}(\hat{r},\hat{t})},\label{edef}\\[0.5ex]
 \hat{q}(\hat{r},\hat{t})&= \frac{\delta^2\,\hat{r}}{\hat{B}(\hat{r},\hat{t})},\label{qdef}\\[0.5ex]
 \hat{I}_p(\hat{r},\hat{t})&= \hat{r}\,\hat{B}(\hat{r},\hat{t}),\label{ipdef}
 \end{align}
 as well as
 \begin{align}
 \frac{d\hat{W}_i}{d\hat{t}} &= \hat{\mit\Gamma}_m(\hat{t}) - \hat{P}_{\rm oh}(\hat{t}),\\[0.5ex]
\frac{d\hat{W}_e}{d\hat{t}} &= \hat{\mit\Gamma}_s(\hat{t}) - \hat{\mit\Gamma}_m(\hat{t}),\\[0.5ex]
\hat{W}_i(\hat{t}) &= \frac{1}{2}\int_0^1\hat{B}^2(\hat{r},\hat{t})\,\hat{r}\,d\hat{r},\\[0.5ex]
  \hat{W}_e(\hat{t}) &= \frac{1}{2}\,l_e\,\hat{I}_p^{\,2}(\hat{t}) - \frac{1}{2}\,\ln[\delta(\hat{t})]\,\hat{I}_p^{\,2}(\hat{t}),\\[0.5ex]
 \hat{\mit\Gamma}_m(\hat{t}) &= \hat{E}(1,\hat{r})\,\hat{I}_p(\hat{t})+\frac{1}{2}\,V(\hat{t})\,\hat{I}_p^{\,2}(\hat{t}), \\[0.5ex]
 \hat{P}_{\rm oh}(\hat{t}) &= \delta^2\int_0^1\skew{3}\hat{T}^{-3/2}(\hat{r},\hat{t})\,\skew{3}\hat{j}^{\,2}(\hat{r},\hat{t})\,\hat{r}\,d\hat{r}.
 \end{align}
 and
 \begin{align}
l_i(\hat{t}) &= \frac{2}{[\hat{r}\,\hat{B}_\theta(\hat{r},\hat{t})]^2_{\hat{r}=1}}\int_0^1\hat{B}_\theta^{\,2}(\hat{r},\hat{t})\,\hat{r}\,d\hat{r},\\[0.5ex]
 \hat{\cal R}_p(\hat{t})&=\frac{\hat{P}_{oh}(\hat{t})}{\hat{I}_p^{\,2}(\hat{t})},\\[0.5ex]
 \hat{V}_{\rm loop}(\hat{t}) &= \frac{\hat{\mit\Gamma}_s(\hat{t})}{\hat{I}_p(\hat{t})}.
\end{align}
and, finally, 
  \begin{align}\label{e174}
\delta^2\,\beta_p\,\left(\frac{\partial\hat{T}}{\partial\hat{t}}-V\,\hat{r}\,\frac{\partial{\hat{T}}}{\partial\hat{r}}\right)&=\frac{1}{\hat{r}}\,\frac{\partial}{\partial\hat{r}}\!\left(\hat{\chi}\,\hat{r}\,\frac{\partial\hat{T}}{\partial \hat{r}}\right) +\delta^2\,\hat{T}^{\,3/2}\,\hat{E}^{\,2} + \delta^2\,\hat{p}_{\rm aux},\\[0.5ex]
\frac{d\hat{W}_{\rm th}}{d\hat{t}} &= \hat{\mit\Gamma}_{\rm th}(\hat{t}) + \hat{P}_{\rm oh}(\hat{t}) + \hat{P}_{\rm aux}(\hat{t}),\\[0.5ex]
\hat{W}_{\rm th}(\hat{t}) &=\frac{3}{2}\,\beta_p\,\delta^2\int_0^1\hat{T}(\hat{r},\hat{t})\,\hat{r}\,d\hat{r},\\[0.5ex]
\hat{P}_{\rm aux}(\hat{t})&= \delta^2\int_0^1\hat{p}_{\rm aux}(\hat{r},\hat{t})\,\hat{r}\,d\hat{r},\\[0.5ex]
  \hat{\mit\Gamma}_{\rm th}(\hat{t})& = \hat{\chi}(1)\left(\hat{r}\,\frac{\partial\hat{T}}{\partial\hat{r}}\right)_{\hat{r}=1} + \frac{3}{2}\,\beta_p\,\delta^2\,V\,(\hat{T})_{\hat{r}=1}.\label{e175}
\end{align}
 Here, $\hat{I}_p(\hat{r},\hat{t})$ is the normalized toroidal plasma current contained within normalized minor radius $\hat{r}$, and $\hat{I}_p(\hat{t})\equiv
 \hat{I}_p(1,\hat{t})$ is the net normalized toroidal current flowing in the plasma. The spatial 
 boundary conditions are (see Sect.~\ref{current}):
 \begin{align}
 \hat{B}(0,\hat{t}) &=0,\label{bc1}\\[0.5ex]
 \hat{B}(1,\hat{t}) &= \hat{I}_p(\hat{t}),\label{bc2}\\[0.5ex]
 \frac{\partial \hat{T}(0,\hat{t})}{\partial\hat{r}}&=0,\label{bc3}\\[0.5ex]
 \hat{T}(1,\hat{t}) & =\zeta\,\hat{T}(0,\hat{t}), \label{bc4}
 \end{align}
 where $\zeta$ is a constant. 
 
\section{Implementation of model}
 
\subsection{Magnetic evolution}
The evolution of the poloidal magnetic field is governed by Eqs.~(\ref{e85}), (\ref{jdef}) and (\ref{edef}), subject to the boundary conditions (\ref{bc1}) and (\ref{bc2}).
The solution of this system of equations, which is a very straightforward task,   is effected by means of the finite-difference approach outlined in Sect.~\ref{appa}. 

\subsection{Temperature evolution}
Let 
\begin{equation}\label{e63}
\hat{\chi}(\hat{r})= f(\alpha)\,(1+\hat{r}^{\,2})^\alpha,
\end{equation}
where
\begin{equation}
f(\alpha) = \frac{1+\alpha}{2^{\,1+\alpha}-1}.
\end{equation}
Note that $\int_0^1\hat{r}\,\hat{\chi}(\hat{r})\,d\hat{r}/\int_0^1\hat{r}\,d\hat{r}=1$, which ensures that $\chi_0$ is the volume averaged perpendicular diffusivity \cite{rf1}.

Let us assume, for the sake of simplicity, that
\begin{equation}
\hat{p}_{\rm aux}(\hat{r},\hat{t})= f_{\rm aux}\,\hat{T}^{\,3/2}(\hat{r},\hat{t})\,\hat{E}^{\,2}(\hat{r},\hat{t}).
\end{equation}
In other words, the auxiliary heating power density has the same profile as the ohmic heating power density, and the net auxiliary heating rate is $f_{\rm aux}$ times the
net ohmic heating rate, where $f_{\rm aux}$ is a constant. 
All of the plasmas considered in this paper are characterized by  $\beta_p\ll 1$ \cite{rf1}. This implies that the thermal energy of the plasma is much less than its magnetic energy.
Neglecting the terms involving $\beta_p$ in Eq.~(\ref{e174}), we can write 
\begin{align}
\frac{dX}{d\hat{r}} &= \hat{r}\,Y^{3/2}(\hat{r}),\\[0.5ex]
\frac{dY}{d\hat{r}}&=- \frac{X(\hat{r})}{\hat{r}\,\hat{\chi}(\hat{r})},
\end{align}
subject to the small-$\hat{r}$ boundary conditions [see Eq.~(\ref{bc3})]
\begin{align}
X(\hat{r})&= \frac{\hat{r}^{\,2}}{2}\,Y_0^{3/2},\\[0.5ex]
Y(\hat{r})&=Y_0- \frac{\hat{r}^{\,2}}{4}\,\frac{Y_0^{3/2}}{\hat{\chi}(0)}.
\end{align}
Here, $Y_0$ is adjusted such that [see Eq.~(\ref{bc4})]
\begin{equation}
Y(1)= \zeta\,Y(0).
\end{equation}
If we assume that $\hat{E}(\hat{r},\hat{t})$ is  spatially uniform then Eqs.~(\ref{jdef})--(\ref{ipdef}), and (\ref{e174})
are satisfied by \cite{rf1}
\begin{align}\label{teq}
\hat{B}(\hat{r},\hat{t}) &= \theta(\hat{t})\,\frac{X(\hat{r})}{\hat{r}},\\[0.5ex]
\skew{3}\hat{j}(\hat{r},\hat{t})&= \theta(\hat{t})\,\delta^{-2}(\hat{t})\,Y^{3/2}(\hat{r}),\\[0.5ex]
\hat{E}(\hat{r},\hat{t}) &= \theta^{-1/5}(\hat{t})\,\delta^{-4/5}(\hat{t})\,(1+f_{\rm aux})^{-3/5},\\[0.5ex]
\hat{q}(\hat{r},\hat{t}) &=\theta^{-1}(\hat{t})\,\delta^2(\hat{t})\,\frac{\hat{r}^{\,2}}{X(\hat{r})},\\[0.5ex]
\hat{I}_p(\hat{r},\hat{t})&= \theta(\hat{t})\,X(\hat{r}),\label{ipx}\\[0.5ex]
\hat{T}(\hat{r},\hat{t}) &= \theta^{4/5}(\hat{t})\,\delta^{-4/5}(\hat{t})\,(1+f_{\rm aux})^{2/5}\,Y(\hat{r}),\label{etemp}
\end{align}
where 
\begin{equation}
\theta(\hat{t}) = \frac{\hat{I}_p(\hat{t})}{X(1)}.
\end{equation}
Note that
\begin{align}
\hat{q}(0,\hat{t}) &= \frac{2}{\theta(\hat{t})\,\delta^2(\hat{t})\,Y_0^{\,3/2}},\\[0.5ex]
\hat{q}(1,0) &=1,
\end{align}
assuming that $\hat{I}_p(0)= 1$. 
In fact, Eqs.~(\ref{teq})--(\ref{ipx})  are only used to set the initial conditions. (In other words, we are only assuming that $E(\hat{r},\hat{t})$ is initially spatially uniform.) 
However, Eq.~(\ref{etemp}) is  employed  to update the electron temperature at each time step, which is a simpler alternative than solving Eq.~(\ref{e174}) at each time step.
Note that if the thermal energy of the plasma is negligible then Eq.~(\ref{eax}) implies that
\begin{equation}\label{e79}
- {\mit\Gamma}_{\rm th}= (1+f_{\rm aux})\,P_{\rm oh}.
 \end{equation}
 In other words, the outward thermal energy flux across the plasma boundary matches the rate at which magnetic energy is converted into thermal
 energy in the plasma due to ohmic heating plus the rate of auxiliary heating. The previous equation can be used as a check that the scaled temperature profile, (\ref{etemp}), is
 consistent with thermal energy balance. 

\subsection{Current ramp}\label{current}
Modern tokamaks are equipped with highly sophisticated control systems that
automatically adjust the currents flowing in the central solenoid and the poloidal field-coils such that a 
 prescribed toroidal plasma current waveform, $\hat{I}_p(\hat{t})$, is generated (within engineering limits) \cite{rf1,romero,ariola,tom}. 
Suppose that the current is ramped  in such a manner that 
 \begin{equation}\label{e61}
 \hat{I}_p(\hat{t}) = F_{\rm ramp}(\hat{t},\hat{t}_0,\hat{t}_I,\hat{\tau}_I,\hat{I}_{p\,1}),
 \end{equation}
 where $F_{\rm ramp}$ is specified in Sect.~\ref{apb}. 
 Thus, the normalized current is ramped down, approximately linearly, from $1$ to $\hat{I}_{p\,1}$,
 between normalized times $\hat{t}_0$ and $\hat{t}_0+\hat{t}_I$. The introduction of the relatively short normalized switch-on time, $\hat{\tau}_I<2\,\hat{t}_I$, ensures that the time derivative of the current is
 continuous in time. 

\subsection{Minor radius ramp}
Current ramp downs are usually accompanied by a simultaneous reduction in the plasma minor radius \cite{muld}. In essence, the plasma is crushed against the X-point. 
Suppose that
\begin{equation}
 \delta(\hat{t}) = \left\{\begin{array}{lcl}
 1 &&\hat{t}_0<\hat{t}\\[0.5ex]
 [F_{\rm ramp}(\hat{t},\hat{t}_0,\hat{t}_a,\hat{\tau}_a,\hat{I}_{p\,1})]^\gamma &&\hat{t}_0<\hat{t}<\hat{t}_0+\hat{t}_a\\[0.5ex]
  [F_{\rm ramp}(\hat{t}_0+\hat{t}_a,\hat{t}_0,\hat{t}_a,\hat{\tau}_a,\hat{I}_{p\,1})]^{\gamma}&& \hat{t}_0+\hat{t}_a<\hat{t}
  \end{array}\right.,
  \end{equation}
  where $\hat{t}_a\leq \hat{t}_I$. 
  Thus, the plasma minor radius is ramped down in such a manner that
  \begin{equation}
  \hat{q}(1,\hat{t}) \simeq[\hat{I}_p(\hat{t})]^{2\,\gamma-1}.
  \end{equation}
  between times 
   $\hat{t}_0$ and
  $\hat{t}_0+ \hat{t}_a$. 
  It follows that 
 \begin{equation}
 V(\hat{t}) = \left\{\begin{array}{lcl}
 0&&\hat{t}_0<\hat{t}\\[0.5ex]
 \gamma\,\dot{F}_{\rm ramp}(\hat{t},\hat{t}_0,\hat{t}_a,\hat{\tau}_a,\hat{I}_{p\,1})/F_{\rm ramp}(\hat{t},\hat{t}_0,\hat{t}_a,\hat{\tau}_a,\hat{I}_{p\,1}) &&\hat{t}_0<\hat{t}<\hat{t}_0+\hat{t}_a\\[0.5ex]
 0&& \hat{t}_0+\hat{t}_a<\hat{t}
  \end{array}\right..
 \end{equation}
The introduction of the relatively short normalized switch-on time, $\hat{\tau}_a<2\,\hat{t}_a$ ensures that $\delta(\hat{t})$ and $V(\hat{t})$ are both continuous in time. 

\section{MHD stability}
\subsection{Cylindrical tearing mode equation}
The MHD stability of a low-$\beta$ tokamak plasma equilibrium is governed by the {\em cylindrical tearing mode equation}\/ \cite{fitz,wesson}:
\begin{equation}
\frac{d^2\psi}{d\hat{r}^{\,2}} + \frac{1}{\hat{r}}\,\frac{d\psi}{d\hat{r}}- \frac{m^2}{\hat{r}^{\,2}}\,\psi
-\frac{(\partial\skew{3}\hat{j}/\partial\hat{r})\,\psi}{\hat{r}\,(1/\hat{q}-1/\hat{q}_s)}=0,
\end{equation}
where $\psi(\hat{r})$ is the perturbed poloidal magnetic flux, $\hat{q}_s= m/(n\,q_a)$, $m>0$ the poloidal mode number of the perturbation, and $n>0$ the toroidal mode number. 

\subsection{Ideal stability}
If there is no resonant surface in the plasma, which means that $\hat{q}(\hat{r})-\hat{q}_s$ does not pass through zero in the region $0<\hat{r}<1$, then the stability of the
plasma to an ideal external-kink mode with $m$ periods in the poloidal direction, and $n$ periods in the toroidal direction, can be assessed as follows. 
We launch a well-behaved solution of the cylindrical tearing mode equation from the magnetic axis ($\hat{r}=0$), and integrate it to the plasma boundary ($\hat{r}=1$). 
We then form the {\em ideal stability index}:
\begin{equation}
{\mit\Delta}_{\rm ideal} = -\left(\frac{d\ln\psi}{d\ln\hat{r}}\right)_{\hat{r}=1}- m\left(\frac{1+1/\hat{r}_w^{\,2\,m}}{1-1/\hat{r}_w^{\,2\,m}}\right).
\end{equation}
Here, $\hat{r}_w(\hat{t}) = \hat{r}_{w\,0}/\delta(\hat{t})$ is the ratio of the minor radius of the vacuum vessel that surrounds the plasma to the instantaneous plasma minor radius,
whereas  $\hat{r}_{w\,0}$ is the ratio at the start of the current ramp. If ${\mit\Delta}_{\rm ideal}>0$ then the Newcomb stability criterion \cite{fried} is violated, and the mode is unstable.
Otherwise, the mode is stable. Generally speaking, the only external-kink modes that are likely to be unstable are ones that are nearly resonant at the plasma boundary \cite{fried}.
Note that external-kink modes evolve on the Alfv\'{e}n time ($10^{-7}$\,s), and are invariably fatal to tokamak discharges. 

\subsection{Tearing stability}
Suppose that there is a resonant surface in the plasma, located at normalized minor radius $\hat{r}_s$. It follows that $\hat{q}(\hat{r}_s)= \hat{q}_s$. Note that the
cylindrical tearing mode equation is singular at the resonant surface. 
In this case, the plasma is potentially unstable to a classical tearing mode that tears and reconnects
magnetic flux at the resonant surface in such a manner as to generate a helical magnetic island chain \cite{fitz,fkr,ruth}.
The intrinsic stability of the tearing mode can be assessed as follows. First, we launch a well-behaved solution of the
cylindrical tearing mode equation from the magnetic axis, and integrate it to the resonant surface. 
Next, we launch a solution from the plasma boundary, such that 
\begin{equation}
\left(\frac{d\ln\psi}{d\ln\hat{r}}\right)_{\hat{r}=1} = - m\left(\frac{1+1/\hat{r}_w^{\,2\,m}}{1-1/\hat{r}_w^{\,2\,m}}\right),
\end{equation}
and integrate it to the resonant surface. We then calculate the {\em tearing stability index}\/ \cite{fkr}:
\begin{equation}
{\mit\Delta}_{\rm tear}=\left(\frac{d\ln\psi}{d\ln\hat{r}}\right)_{\hat{r}=\hat{r}_{s+}} - \left(\frac{d\ln\psi}{d\ln\hat{r}}\right)_{\hat{r}=\hat{r}_{s-}}.
\end{equation}
In principle, a classical tearing mode with $m$ periods in the poloidal direction, and $n$ periods in the toroidal direction, is unstable when ${\mit\Delta}_{\rm tear}>0$, and
stable otherwise. 

However, this is not the end of the story. Tokamak plasmas possess stabilizing average magnetic field-line curvature that renders them somewhat resilient to tearing modes. 
In fact, the tearing stability index has to exceed a certain critical value before the mode becomes unstable \cite{ggj,ggj1}. The critical stability index
takes the following form \cite{lut,rf2}: 
\begin{equation}
{\mit\Delta}_{\rm crit} =\frac{ \sqrt{2}\,\pi^{3/2}\,D_R}{\hat{\delta}_d},
\end{equation}
where
\begin{equation}
D_R = \left[\left(\frac{2\,q^2}{s^2}\,\hat{r}\,\frac{dP}{d\hat{r}}\right)\left(1-\frac{1}{q^2}\right)\right]_{\hat{r}=\hat{r}_s},
\end{equation}
and $s=\ln\hat{q}/d\ln\hat{r}$, $P= 2\,\mu_0\,n_e\,T_e/B_0^{\,2}$ (here, we are assuming that the electron and ion temperatures are equal), with
\begin{align}
\hat{\delta}_d &=\sqrt{8}\left(\frac{\chi_\perp}{\chi_\parallel}\right)^{1/4}\,\frac{1}{[\hat{r}\,\delta(\hat{t})\,s\,n\,a/R_0]^{1/2}},\\[0.5ex]
\chi_{\parallel} &= \frac{\chi_\parallel^{\rm smfp}\,\chi_{\parallel}^{\rm lmfp}}{\chi_{\parallel}^{\rm smfp}+ \chi_{\parallel}^{\rm lmfp}},\\[0.5ex]
\chi_{\parallel}^{\rm smfp} &= \frac{1.581\,\tau_{ee}\,v_{t\,e}^{\,2}}{1+0.2535\,Z},\\[0.5ex]
\chi_{\parallel}^{\rm lmfp }&= \frac{2\,R_0\,v_{t\,e}}{\pi^{1/2}\,n\,s\,\hat{\delta}_d},\\[0.5ex]
v_{t\,e} &= \left(\frac{2\,T_e}{m_e}\right)^{1/2},\\[0.5ex]
\tau_{ee} &= \frac{6\sqrt{2}\,\pi^{3/2}\,m_e^{1/2}\,T_e^{\,3/2}}{\ln{\mit\Lambda}\,e^4\,c^4\,\mu_0^{\,2}\,n_e}.
\end{align}
All quantities in the previous six equations are evaluated at the resonant surface. We conclude that the tearing mode is only unstable when
the {\em effective tearing stability index},
\begin{equation}
{\mit\Delta}_{\rm eff} = {\mit\Delta}_{\rm tear}-{\mit\Delta}_{\rm crit}, 
\end{equation}
is positive. The model outlined in the previous eight equations is a modification of the original calculation of Glasser, Greene \& Johnson\,\cite{ggj,ggj1} that takes the anomalously large
perpendicular diffusivity in tokamak plasmas into account \cite{rf2}. This model was found to give good agreement with the XTOR toroidal plasma stability code \cite{lut}. 
Note that it makes sense to include curvature stabilization, while neglecting the destabilizing effect of the perturbed bootstrap current, when calculating the
effective linear tearing stability index \cite{car}, because the former is a linear
effect whereas the latter only appears at finite mode amplitude \cite{rf3}. 

\begin{figure}
\centerline{\includegraphics[width=\textwidth]{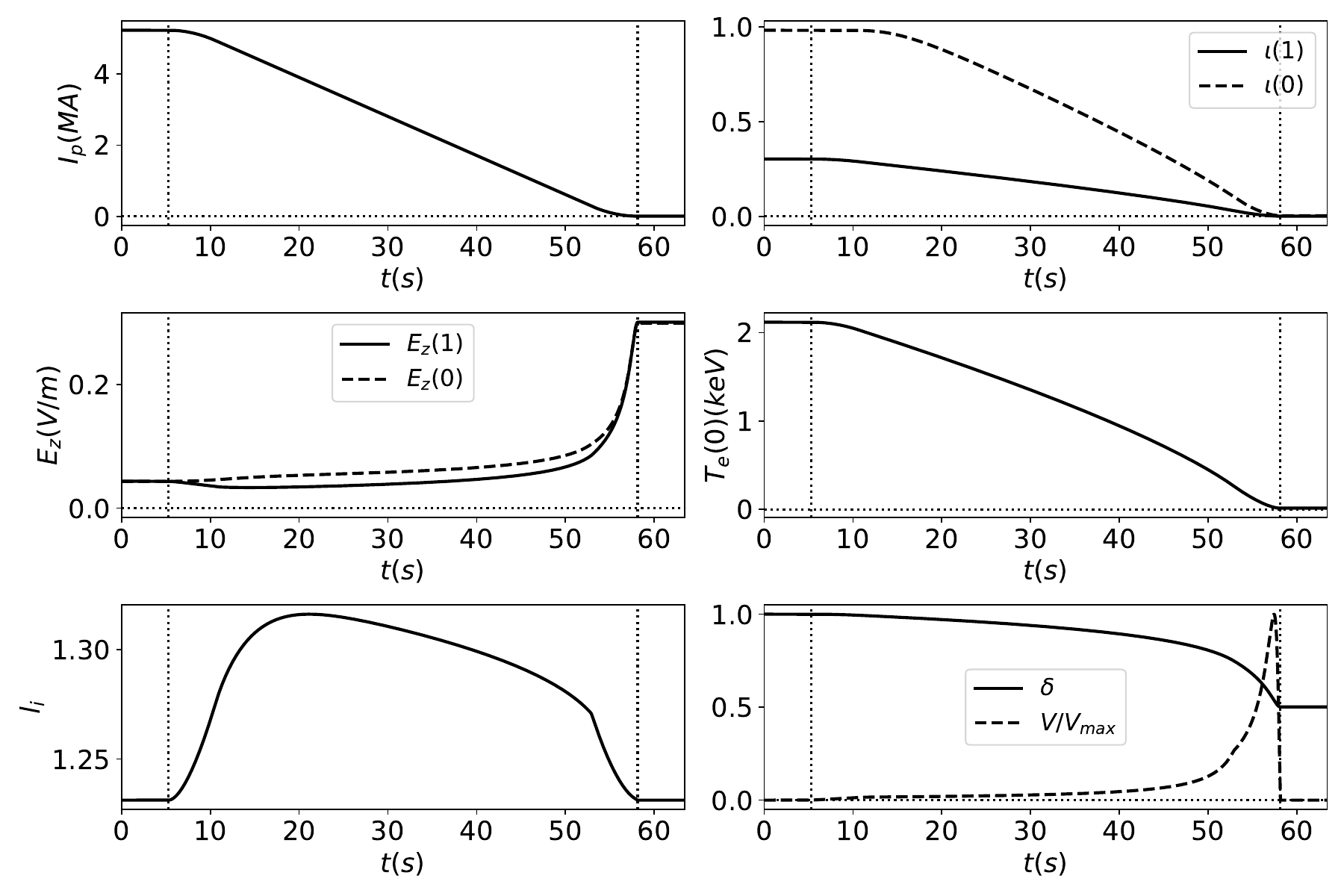}}
\caption{Overview of Simulation 1. The following data is plotted versus time: (top left) plasma current, $I_p$; (top right) central rotational transform, $\iota(0)$, and
edge rotational transform, $\iota(1)$; (middle left) central electric field, $E_z(0)$, and edge electric field,  $E_z(1)$; (middle right) central electron temperature, $T_e(0)$;
(bottom left) normalized internal self-inductance, $l_i$; (bottom right) relative minor radius $\delta$, and $V\equiv d\ln\delta/d\hat{t}$.}\label{fig1}
\end{figure}

Even this is not the end of the story, because, unlike external-kink modes, tearing modes are not invariably fatal to tokamak discharges. Indeed,
some tearing modes saturate in such a manner that they impair the energy confinement properties of the plasma without triggering a disruption \cite{white}. 
The saturated radial width, $\hat{W}_{\rm sat}$, of the magnetic island chain associated with a tearing mode is given by \cite{fitz,has}: 
\begin{equation}
\hat{W}_{\rm sat} = \frac{{\mit\Delta}_{\rm eff}\,\hat{r}_s}{0.8\,\alpha_s^{\,2} -0.27\,\beta_s-0.09\,\alpha_s},
\end{equation}
where 
\begin{align}
\alpha_s &=- \left(\frac{\hat{q}\,\hat{r}\,\partial\skew{3}\hat{j}/\partial \hat{r}}{s}\right)_{\hat{r}=\hat{r}_s},\\[0.5ex]
\beta_s &=- \left(\frac{\hat{q}\,\hat{r}^{\,2}\,\partial^2\skew{3}\hat{j}/\partial^2 \hat{r}}{s}\right)_{\hat{r}=\hat{r}_s}.
\end{align}
Here, $\hat{W}_{\rm sat}$ is measured with respect to the instantaneous minor radius of the plasma. The neglect of the perturbed bootstrap
current at finite mode amplitude can be justified because the comparatively cold plasmas that occur during the current ramp-down phase 
are characterized by $\beta_p\ll 1$.

\subsection{Mode locking}
We need a criterion that will allow us to distinguish between a benign tearing mode, and one that is likely to trigger a disruption. 
Now, tearing modes are convected by the electron fluid at the associated resonant surface, which implies that they are born rotating in the laboratory
frame \cite{rf4}. However, the magnetic perturbation associated with a rotating tearing mode excites eddy currents in the vacuum vessel that surrounds the plasma,
and these current lead to the development of an electromagnetic breaking  torque acting at the resonant surface that  slows the rotation of the electron fluid, and, hence, of the mode \cite{nave}. Above a certain critical mode amplitude, the rotation of the  mode is suddenly arrested (because it is reduced to such a low value that the mode
can lock to a stray error-field), and the mode
becomes a so-called {\em locked mode}\/ \cite{rf4,lh,rf5}. Now, there is a very strong experimental correlation between mode locking and disruptions \cite{sweeney, jet}. 
Hence, in this paper, we shall assume that a tearing mode whose amplitude is not large enough to lock to the vacuum vessel is benign, whereas a tearing mode
that locks to the vessel triggers a disruption. 

Our mode locking model  is a simplified version of the model described  in Chap.~10 of Ref.~\cite{fitz1}, 
which was based on one first presented in Ref.~\cite{rf4}. 
In this model, mode locking is triggered by a breakdown  of the
balance between the electromagnetic locking torque and the viscous restoring torque that occurs when the mode frequency has been reduced to half
of its original value. 
The model predicts that the critical value of $\hat{W}_{\rm sat}$
at which mode locking occurs is 
\begin{equation}
\hat{W}_{\rm crit} =\frac{4\,\hat{r}_s\,(\omega_{\ast\,e}\,\tau_H)^{1/2}}{E_{sw}^{\,1/2}}\left(\frac{\tau_w}{\tau_V}\right)^{1/4}\left(\frac{q_s}{\epsilon_s}\right)^{1/2},
\end{equation}
where
\begin{align}
\tau_H &= 4.5\times 10^{-7}\,\frac{R_0\,(n_{20}\,M)^{1/2}}{s(\hat{r}_s)\,n\,B_0},\\[0.5ex]
\tau_V& = \frac{1}{2}\,\ln\left(\frac{1}{\delta\,\hat{r}_s}\right)\frac{\delta^2\,\hat{r}_s^{\,2}}{\chi_0},\\[0.5ex]
\omega_{\ast\,e} &= - 10^3\,\frac{m}{\delta^{\,2}\,\hat{r}_s}\,\frac{T_0}{B_0}\left(\frac{\partial\hat{T}}{\partial\hat{r}}\right)_{\hat{r}=\hat{r}_s},\\[0.5ex]
E_{sw} &=\frac{2\,m\,(\delta\,\hat{r}_s/\hat{r}_{w\,0})^m}{1- (\delta\,\hat{r}_s/\hat{r}_{w\,0})^{2m}},
\end{align}
$\epsilon_s = \hat{r}_s\,\delta\,a/R_0$, and $q_s=m/n$. 
Here, $M$ is the ion mass number, $\tau_H$ the hydromagnetic timescale, $\tau_V$ the viscous restoring time, $\tau_w$ the L/R time of the vacuum vessel (which is defined
in Ref.~\cite{rf5}),  and $\omega_{\ast\,e}$ the electron diamagnetic frequency.
In writing the previous formulae, we have made a number of simplifying approximations.
First, we have assumed that the ion fluid is initially at rest, which implies that the electron fluid rotates at twice the electron diamagnetic velocity. Second, we  have assumed that the perpendicular toroidal momentum diffusivity takes the value $\chi_0$. Third, we have assumed that poloidal flow damping
cancels out the poloidal component of the electromagnetic torque at the resonant surface \cite{rf4,nave}. Fourth, we have assumed that the vacuum vessel can be treated as
a thin shell. Finally, we have calculated the plasma-vessel coupling
coefficient, $E_{sw}$, while neglecting any equilibrium plasma currents outside the resonant surface. 
\section{Simulations}

\subsection{ITER}
We shall use our model to simulate  controlled current ramp downs in the ITER tokamak. Our simulations all utilize the following set of common parameters:
$R_0=6.2\,{\rm m}$, $a=2.0\,{\rm m}$, $B_0=5.3\,{\rm T}$, 
$n_e=1\times 10^{20}\,{\rm m}^{-3}$, $M=2.5$, $\chi_0=1\,{\rm m^2/s}$, $Z=4$,  
$\hat{r}_{w\,0}=1.2$, and $\tau_{w}=23\,{\rm ms}$   \cite{creely,iter1,rf5}. Here, the rather large value of $Z$ is intended to compensate for the
neglect of the neoclassical enhancement of plasma resistivity. We also use the following common numerical parameters: $\zeta=0.01$, $I=500$,  and  $D=1.0$.
(Sse Sect.~\ref{appa}.)

\begin{figure}
\centerline{\includegraphics[width=0.8\textwidth]{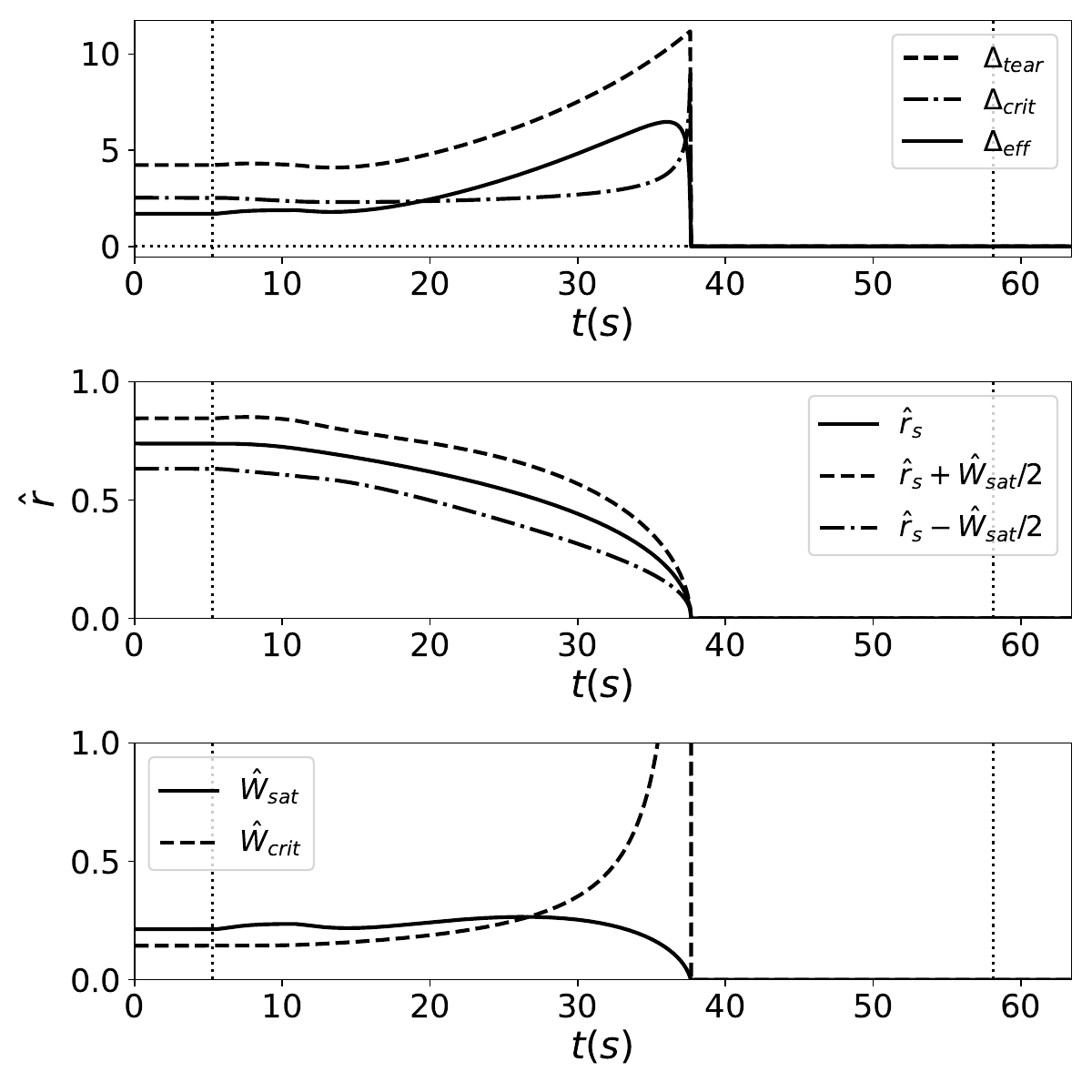}}
\caption{Stability of $m=2/n=1$ tearing mode during Simulation 1. Here, ${\mit\Delta}_{\rm tear}$ is the tearing stability index, ${\mit\Delta}_{\rm crit}$ the critical value of the index that must be exceeded before the mode grows, ${\mit\Delta}_{\rm eff}\equiv {\mit\Delta}_{\rm tear}-{\mit\Delta}_{\rm crit}$. Moreover,
$\hat{r}_s$ is the resonant surface radius, and $\hat{W}_{\rm sat}$ the saturated island width, and $\hat{W}_{\rm crit}$ the critical island width for mode locking.}\label{fig2}
\end{figure}

\subsection{Simulation 1}
Simulation 1 employs the following additional parameters: $\alpha=0.0$ [i.e., a uniform $\chi_\perp(r)$ profile], $f_{\rm aux}=0.0$ 
(i.e., no auxiliary heating), and $q_a=3.3$. The value of $q_a$ is chosen such that the central safety-factor lies just above unity at the start of the current ramp. 
The scale quantities in the simulation take the following values: $T_0= 1.03\,{\rm keV}$, $\tau_R= 52.9\,{\rm s}$, $\beta_p= 7.63\times 10^{-2}$, $I_0=5.23\,{\rm MA}$, $E_0=1.98\times 10^{-2}\,{\rm V/m}$, $W_0=0.213\,{\rm GJ}$, ${\cal R}_0=1.47\times 10^{-7}\,\Omega$, and $V_0= 0.770\,{\rm V}$. The rather low value of $I_0$ (compared to ITER's actual toroidal
current, which is 15\,MA) is due to the fact that the cylindrical relation between the plasma current and the edge safety-factor, 
$I_p= 2\pi\,a^2\,B_0/(\mu_0\,R_0\,q_a)$,  does not take into account the fact that ITER plasmas will be both vertically elongated and triangular \cite{uckam}.
The remaining simulation parameters are $\hat{I}_{p\,1}=10^{-3}$, $\hat{t}_0= 0.1$. 
$\hat{t}_I=\hat{t}_a= 1.0$, $\hat{\tau}_I=\hat{\tau}_a=0.1$,  and $\gamma=0.1$. 

Figure~\ref{fig1} gives an overview of the current ramp down in Simulation 1. It can be seen that the current is ramped down, approximately linearly in time, in
about 50 s, while the plasma minor radius is simultaneous reduced by a factor of 2. Roughly speaking, this is what is envisaged in an actual ITER current ramp down
\cite{imb,vries,muld}. Note that the edge rotational transform (which is the inverse of the safety-factor) ramps down almost linearly in time. 
However, there is a delay of about 10 s before the central rotational transform starts to ramp down. The central and edge electric fields remain almost
the same as one another throughout the ramp. This implies that the electric field remains approximately spatially uniform in the plasma during the ramp.
In fact, the ramp has almost identical properties to the self-similar ramps described in Ref.~\cite{rf1}. Finally, there is a very modest increase in the
normalized plasma self-inductance during the ramp.

The next question that we need to ask is whether the plasma remains MHD stable during the current ramp down. This question  was not addressed in  perviously  published, 
and much more sophisticated,  studies of current ramp down in tokamak plasmas \cite{asdex, demo, muld}, and is the main question addressed, in an albeit highly simplified
fashion, in this paper.  It turns out that all external-kink modes are stable during the ramp. Furthermore, all tearing mode, other than the $m=2/n=1$ mode, are stable
during the ramp. Figure~\ref{fig2} illustrates the stability of the 2/1 tearing mode during the ramp. It can be seen that the mode is initially unstable; in other words, 
${\mit\Delta}_{\rm eff}>0$ at $\hat{t}=0$. As the current is ramped down, the effective tearing stability rises. However,  the radius of the resonant surface decreases, until the
surface is expelled from the plasma, via the magnetic axis, about two-thirds of the way through the ramp. The initial saturated island width is about 20\% of the plasma
minor radius, but actually decreases as the resonant surface moves inward. Nevertheless, the initial island width exceeds the critical
width above which mode locking is triggered. Hence, we deduce that the plasma in Simulation 1 would probably have  disrupted prior to the initiation of the current
ramp.

\begin{figure}
\centerline{\includegraphics[width=\textwidth]{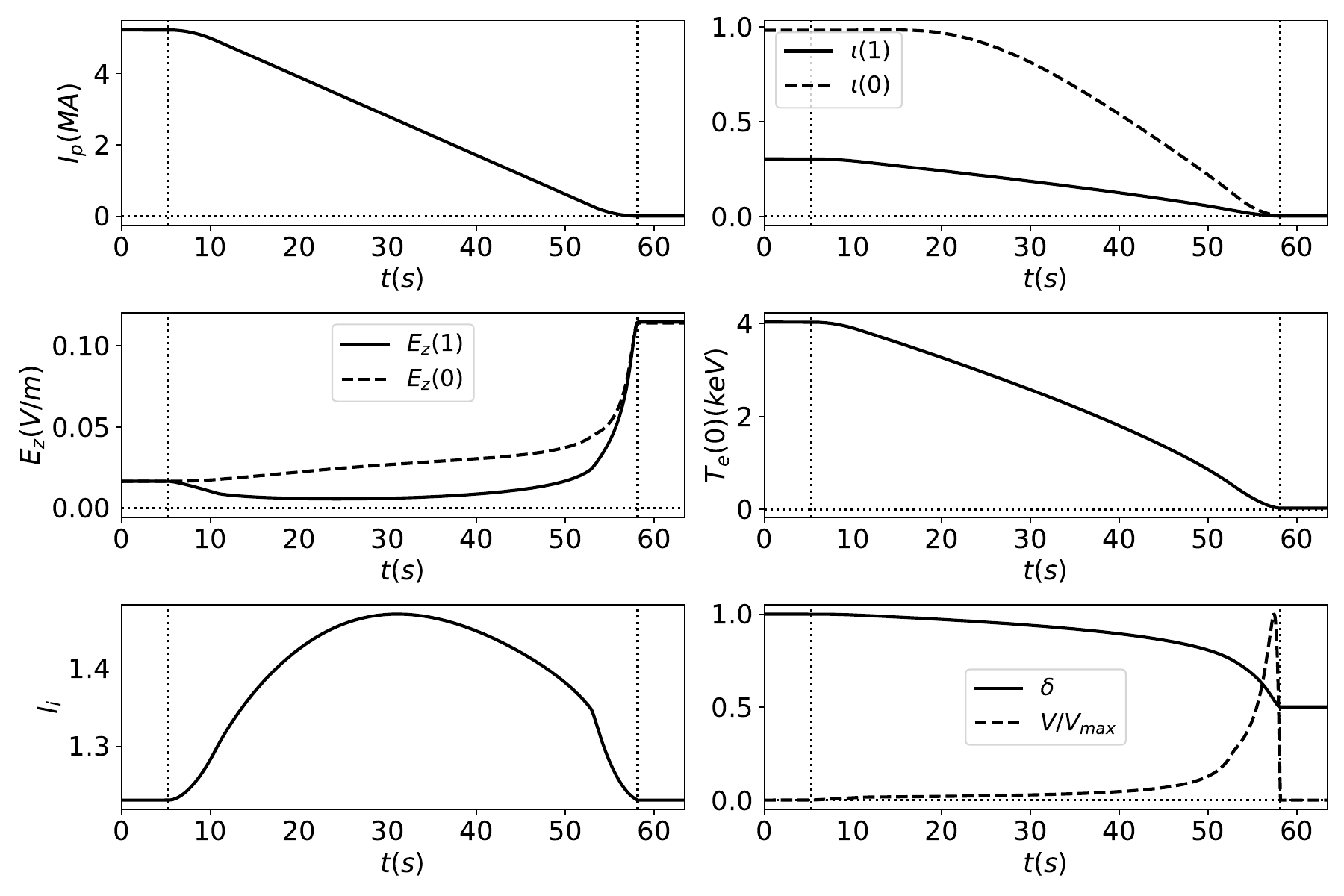}}
\caption{Overview of Simulation 2. See caption to Fig.~\ref{fig1}}\label{fig3}
\end{figure}

The main problem with Simulation 1 is the fact that the 2/1 tearing mode is already unstable before the current ramp down commences. This is not realistic. 
While the 2/1 classical tearing mode was ubiquitous in early small tokamaks, it is fairly uncommon in modern large tokamaks. The reason for this is that large
tokamaks feature hotter plasmas than small tokamaks, and the mode is stabilized by favorable average magnetic curvature \cite{fitz1,ggj,rf2}. This stabilizing effect is
proportional to the plasma pressure, and, hence, to the plasma temperature in tokamaks with similar electron number densities. We conclude that the problem
with Simulation 1 is that the plasma is initially too cold. The reason that the plasma is cold is that it is only subject to ohmic heating. However, this is not realistic.
It is envisaged that the ITER current ramp down will commence when the plasma is still subject to $\alpha$ particle heating from nuclear fusion reactions \cite{muld}. 
Moreover, the plasma will  be in the H-mode phase at the start of the ramp \cite{muld}. Both of these facts point to the presence of plasma heating, 
in addition to ohmic heating, at the
start of ramp. 

\subsection{Simulation 2}
Simulation 2 is identical to Simulation 1, except that $f_{\rm aux}=4$. In other words, the additional plasma heating is four times the ohmic heating. The additional
heating decays in time at the same rate as the ohmic heating. This represents the gradual quenching of $\alpha$ particle heating, and a ramp down of
any auxiliary heating. As illustrated in Fig.~\ref{fig3}, the current ramp in Simulation 2 is very similar to that shown in Fig.~\ref{fig1}. The main differences are that the initial temperature is about 4 keV, there is a greater difference between the electric field at the plasma boundary and that at
the magnetic axis, and the increase in $l_i$ during the ramp is slightly larger. Incidentally, the thermal energy balance relation (\ref{e79}) is found to hold almost exactly in
Simulations 1 and 2, indicating that the scaled temperature profile, (\ref{etemp}), is consistent with thermal energy balance. 

\begin{figure}
\centerline{\includegraphics[width=0.8\textwidth]{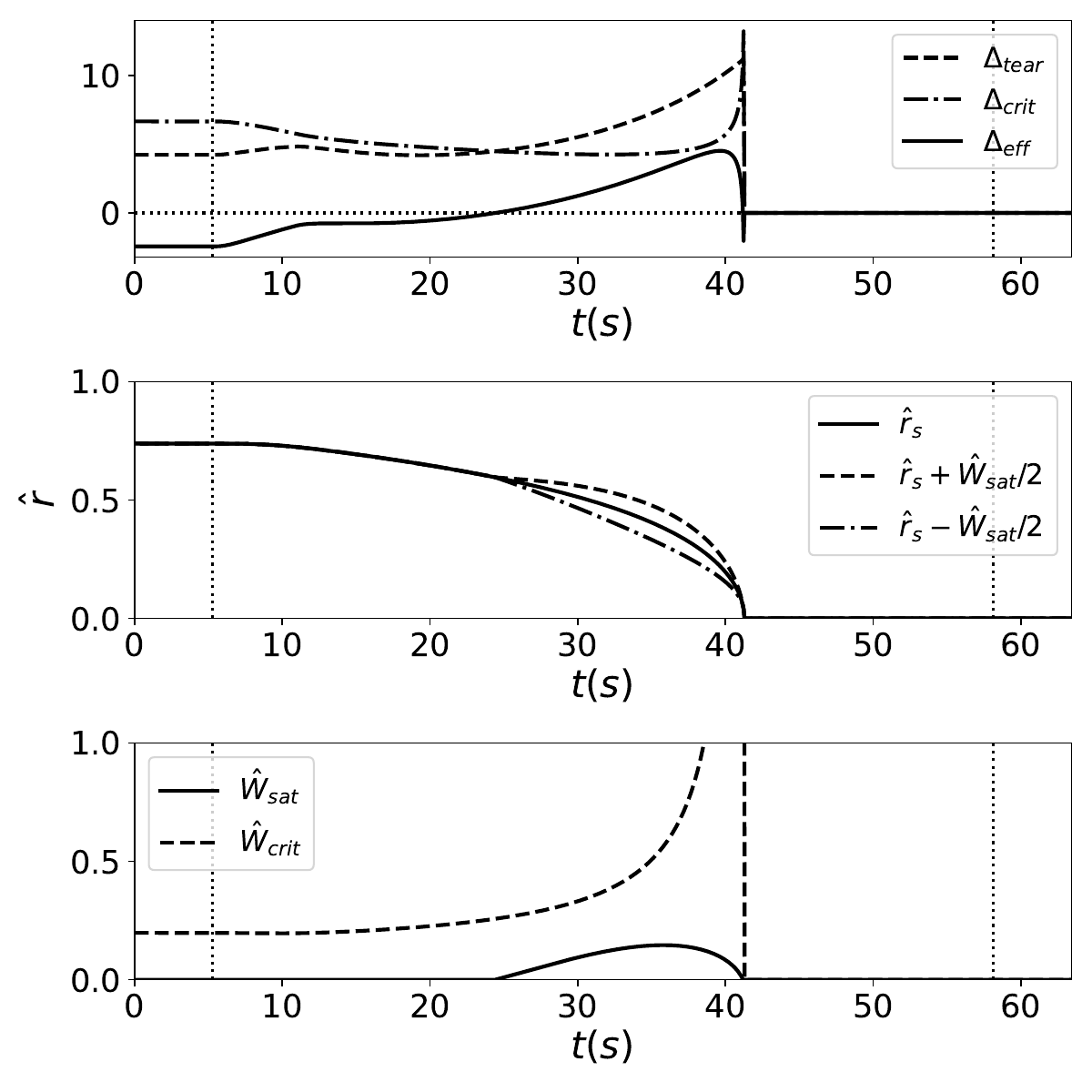}}
\caption{Stability of $m=2/n=1$ tearing mode during Simulation 2. See caption to Fig.~\ref{fig2}. }\label{fig4}
\end{figure}

As before, all external-kink modes, and all tearing modes other than the 2/1 mode, are stable throughout the current ramp. Figure~\ref{fig4} illustrates the stability of
the 2/1 tearing mode during the  ramp. It can be seen that the tearing mode is initially stable. This is because ${\mit\Delta}_{\rm crit}$ is higher in
Simulation 2 than in Simulation 1, since the plasma is hotter in the former case. The mode becomes unstable about halfway through the ramp. However, at
this stage, the resonant surface is deep within the plasma. Moreover, the saturated island width is fairly small, and remains well below the level needed to trigger mode locking.
  Hence, we deduce that the 2/1 tearing mode in Simulation 2 is benign in nature, and is very unlikely to trigger a disruption. 

We can draw two conclusions from Simulation 2. First, the envisaged ITER current ramp-down scenario, in which the current is ramped down in about
60 s, and the plasma minor radius is simultaneously decreased by a factor of 2 \cite{imb,vries,muld}, seems perfectly feasible. The main obstacle to the
ramp is the 2/1 tearing mode. However, this obstacle can be overcome by ensuring that the plasma is sufficiently hot at the start of the ramp. 
The second conclusion is that reliance on $q_a$-$l_i$ diagrams to assess MHD stability can give rise to misleading conclusions. The
plasmas in Simulations 1 and 2 are initially in almost exactly the same place in $q_a$-$l_i$ space. However, the former is MHD unstable, whereas the
latter is stable. 

\begin{figure}
\centerline{\includegraphics[width=\textwidth]{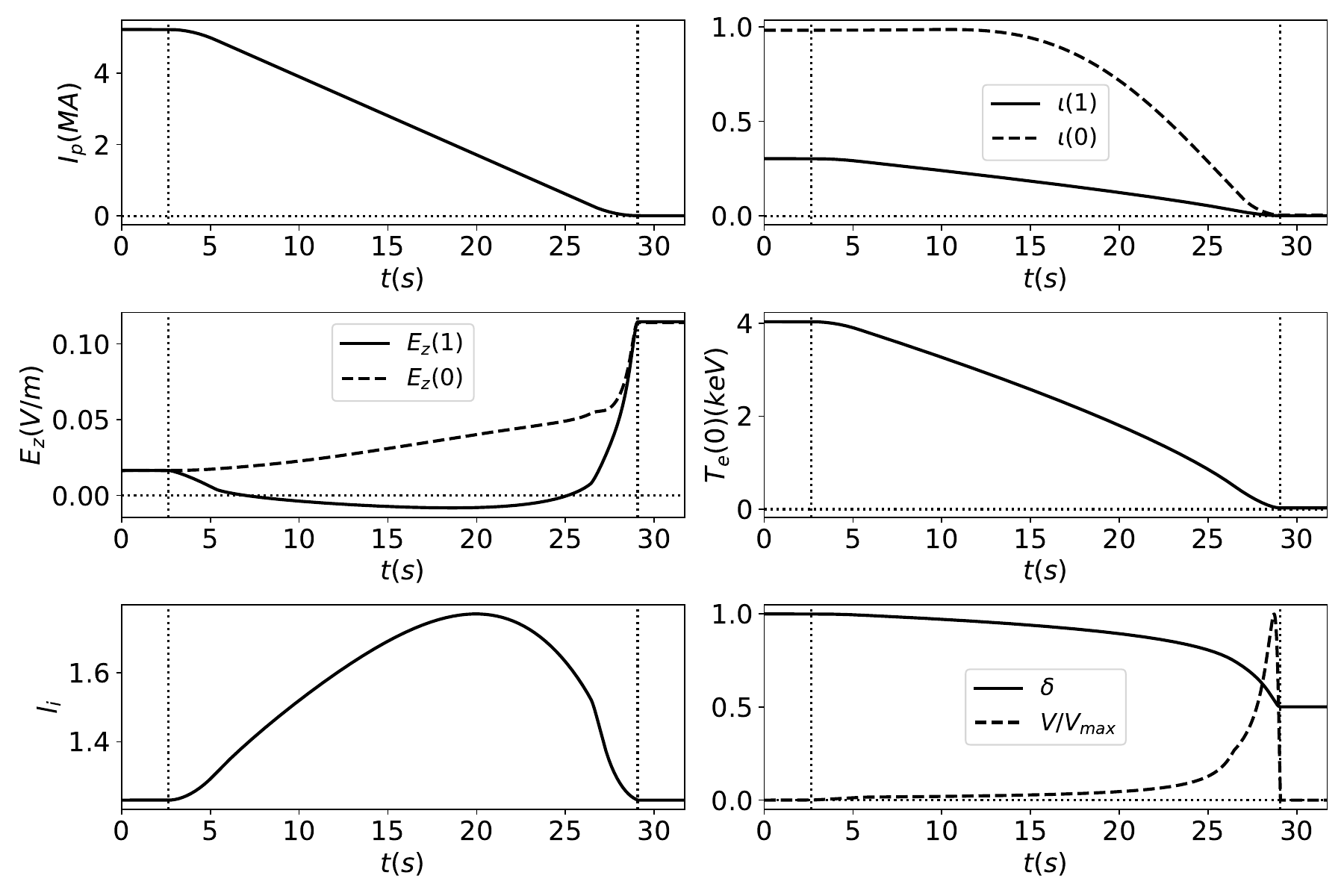}}
\caption{Overview of Simulation 3. See caption to Fig.~\ref{fig1}}\label{fig5}
\end{figure}

\subsection{Simulation 3}
Simulation 3 is the same as Simulation 2, except that the plasma current and minor radius are ramped down twice as fast in the former case. 
As illustrated in Fig.~\ref{fig5}, the rapid ramp down of the plasma current causes the electric field at the plasma boundary to become negative, indicating
that the electric field acts to drive a reversed current density in the outer regions of the plasma \cite{muld}. The increase in $l_i$ during the ramp is
also significantly larger than that shown in Fig.~\ref{fig3}. 

\begin{figure}
\centerline{\includegraphics[width=0.8\textwidth]{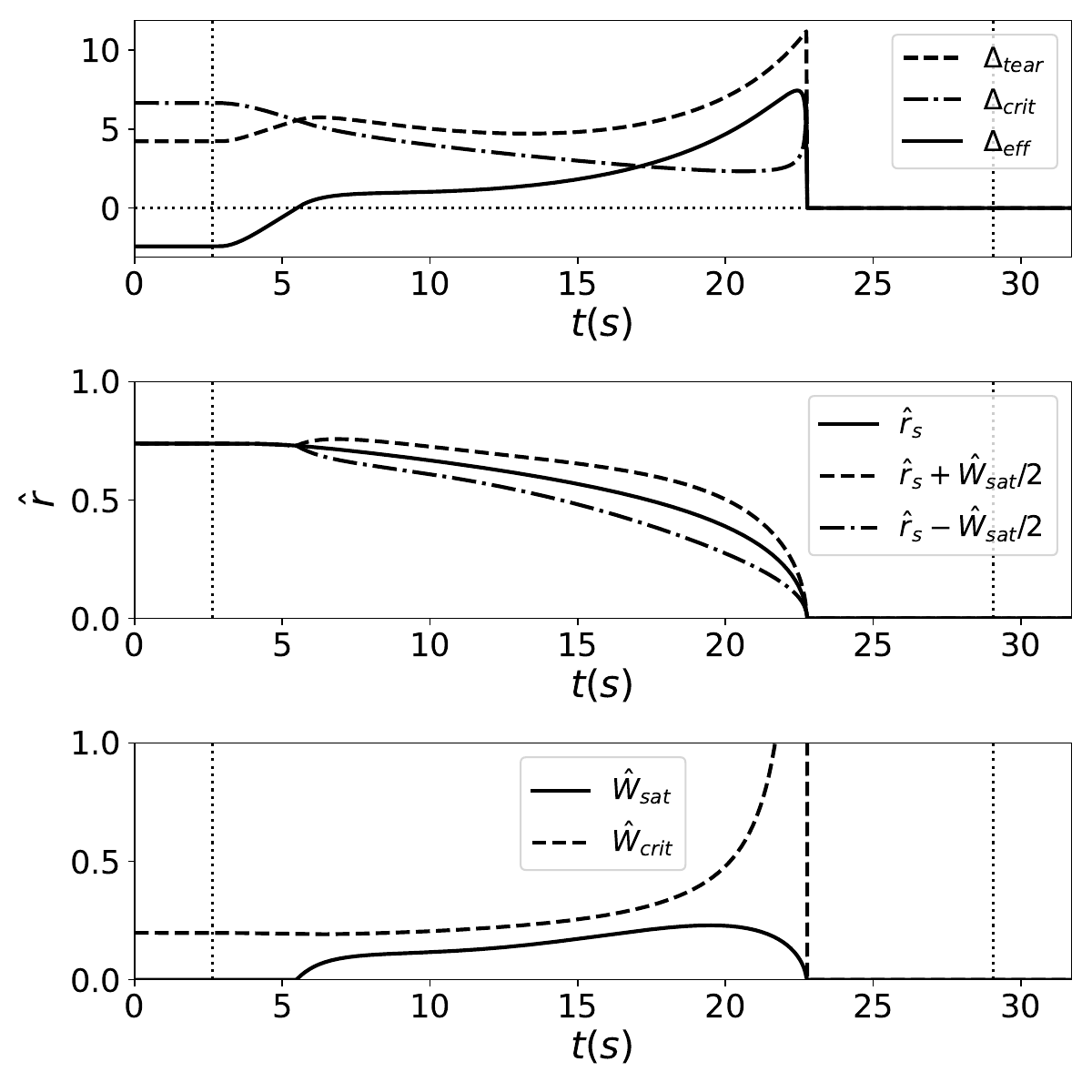}}
\caption{Stability of $m=2/n=1$ tearing mode during Simulation 3. See caption to Fig.~\ref{fig2}. }\label{fig6}
\end{figure}

As before, all external-kink modes, and all tearing modes other than the 2/1 mode, are stable throughout the current ramp. 
Figure~\ref{fig6} illustrates the stability of the 2/1 tearing mode during the  ramp. It can be seen that the mode is driven unstable during the ramp. Moreover, the
saturated island width is significantly larger than that shown in Fig.~\ref{fig4}. Although the island width does not attain its maximum value until the resonant surface
has shifted inward by a significant amount, it comes uncomfortably close to the mode locking threshold. We conclude that the 2/1 tearing mode in Simulation 3 
is probably benign. 

\subsection{Simulation 4}
\begin{figure}
\centerline{\includegraphics[width=\textwidth]{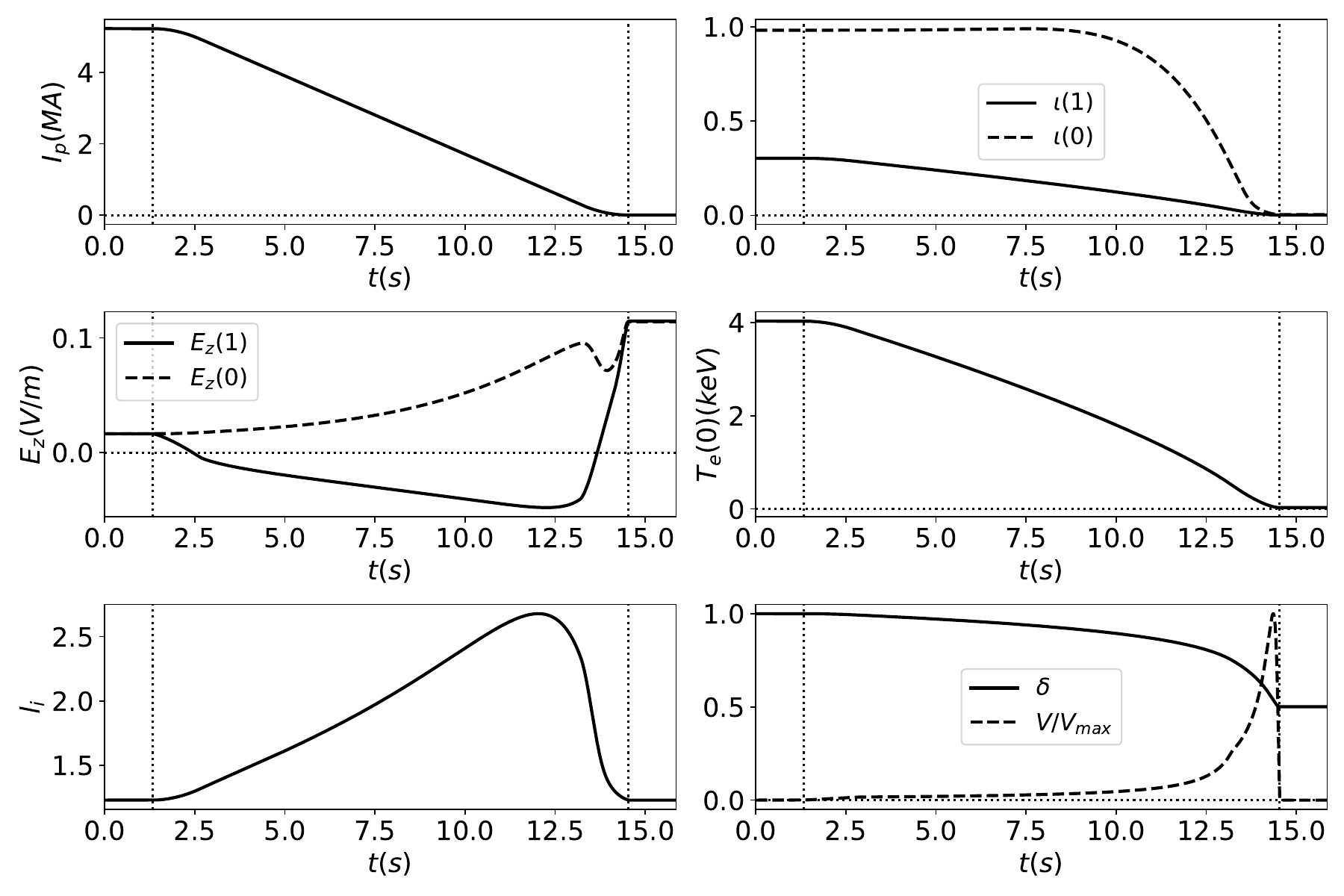}}
\caption{Overview of Simulation 4. See caption to Fig.~\ref{fig1}}\label{fig7}
\end{figure}

Simulation 4 is the same as Simulation 3, except that the plasma current and minor radius are ramped down twice as fast in the former case. 
As illustrated in Fig.~\ref{fig7}, the very rapid ramp down of the plasma current causes the electric field at the plasma boundary to become extremely negative, indicating
that the electric field acts to drive a strong reversed current density in the outer regions of the plasma \cite{muld}. The increase in $l_i$ during the ramp is
also quite marked. 

\begin{figure}
\centerline{\includegraphics[width=0.8\textwidth]{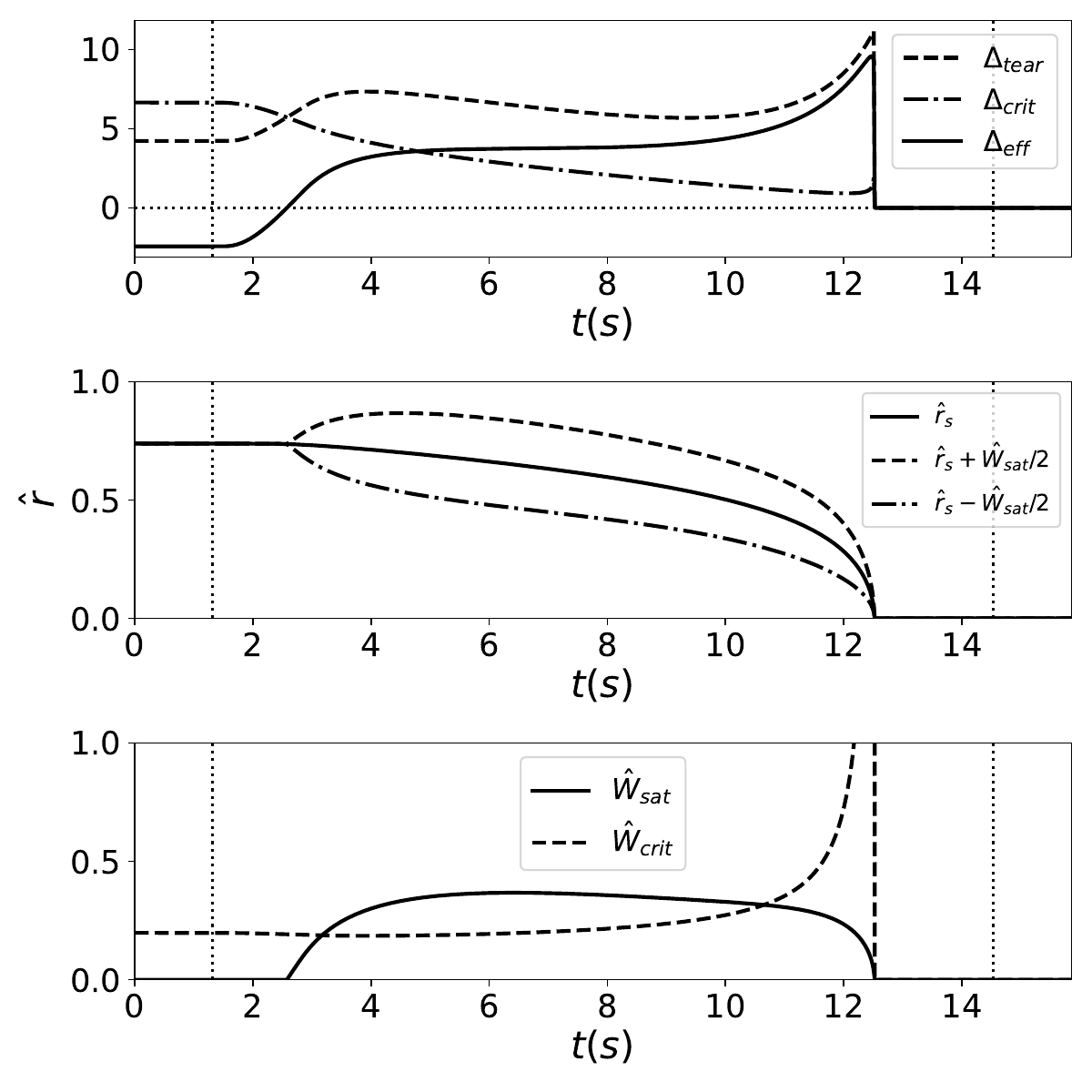}}
\caption{Stability of $m=2/n=1$ tearing mode during Simulation 4. See caption to Fig.~\ref{fig2}. }\label{fig8}
\end{figure}

As before, all external-kink modes, and all tearing modes other than the 2/1 mode, are stable throughout the current ramp. 
Figure~\ref{fig8} illustrates the stability of the 2/1 tearing mode during the  ramp. It can be seen that the mode is driven unstable during the ramp. Moreover, the
saturated island width is very large, and quickly exceeds the critical width for mode locking. 
 Thus, in this case, it is almost certain that the island chain would lock to the wall and trigger a disruption. 

It is interesting to observe, from a comparison of Figs.~\ref{fig1}--\ref{fig8}, that while larger peak $l_i$ values correspond to larger
peak 2/1 island widths, the times at which the peak $l_i$ values and peak island widths are attained do not seem to be correlated. Note that a $\hat{\chi}(\hat{r})$
profile that increases with minor radius [i.e., $\alpha>0$ in Eq.~(\ref{e63})] alleviates the problem of the 2/1 tearing mode to some extent, because it
increases $q_a$, and, thus, pushes the 2/1 resonant surface farther inside the plasma. 

\subsection{$q_a$-$l_i$ diagram}
\begin{figure}
\centerline{\includegraphics[width=0.8\textwidth]{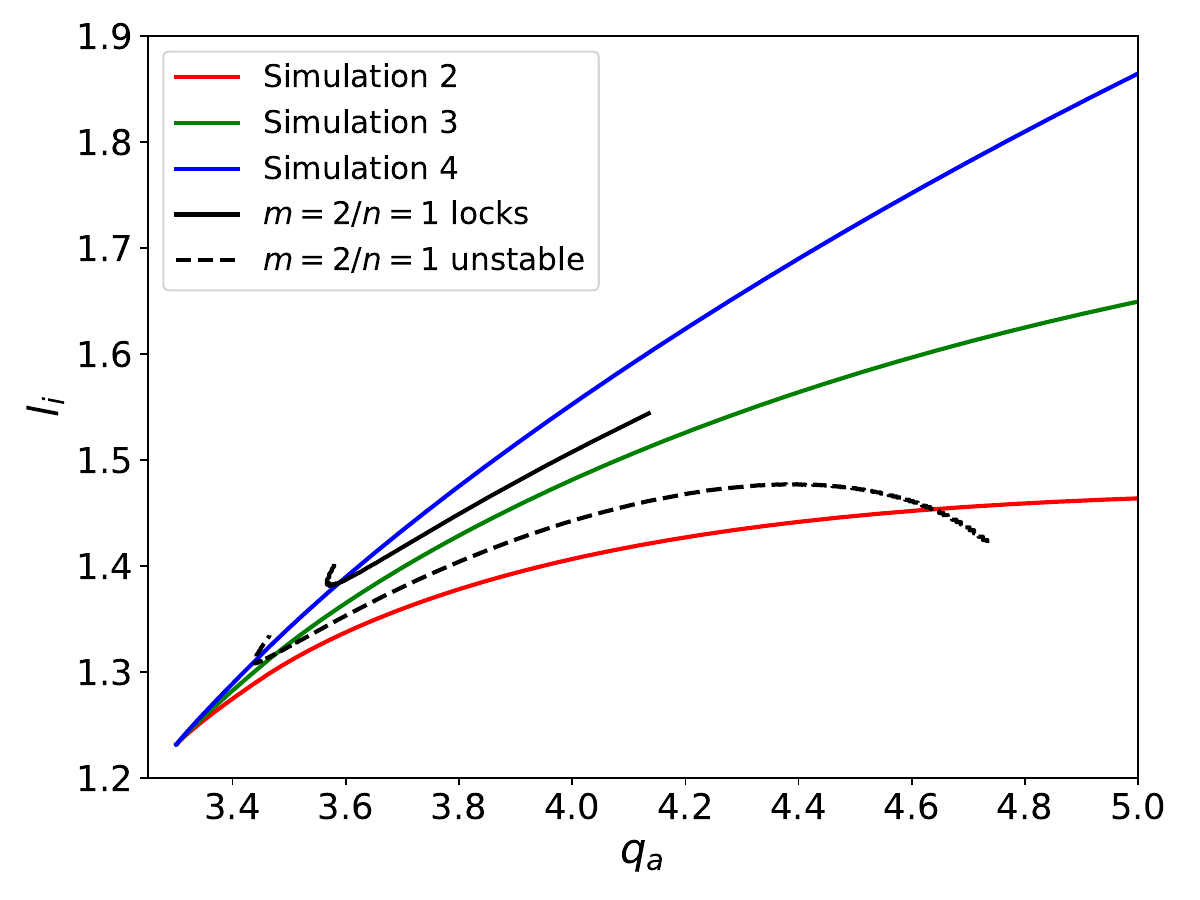}}
\caption{Trajectories of Simulations 2, 3, and 4 in $q_a$-$l_i$ space. The dashed black line shows the $m=2/n=1$ tearing mode stability limit,
whereas the solid black line shows the $m=2/n=1$ locking threshold. }\label{fig9}
\end{figure}

Although we have expressed some reservations about $q_a$-$l_i$ diagrams, it seems reasonable to use one to compare current ramp downs performed at different
ramp rates, starting from the same initial plasma state. 
Figure~\ref{fig9} shows the trajectories of Simulations 2, 3, and 4 in $q_a$-$l_i$ space. (Here, $q_a$ is the instantaneous safety-factor value at the plasma boundary.)  Also shown is the stability limit, as well as the mode locking threshold, for the
$m=2/n=1$ tearing mode. 
 The limit and threshold are 
obtained by  running many simulations that are similar to Simulations 2, 3, and 4, and that employ a range of different current ramp down rates. If a trajectory in Fig.~\ref{fig9} crosses the
stability limit, or the locking threshold, from below to above, then instability and mode locking occur,  respectively.  In the latter case, we infer that a disruption is triggered. It is clear from the figure that, while Simulation 2 stays well away from the
mode locking threshold, Simulation 2 skirts the threshold, and Simulation 3 crosses the threshold. 

\section{Summary and discussion}
We have simulated  the controlled  ramp down of the toroidal plasma current in the ITER tokamak using a simple mode that employs cylindrical geometry. We have simultaneously calculated the MHD stability of the plasma throughout the whole ramp. 
We find that the only potentially unstable MHD mode is the $m=2/n=1$ classical tearing mode. The envisioned 60 s ramp down of the plasma current in ITER is
determined to be perfectly feasible, provided that the plasma is sufficiently hot at the start of the ramp. However, attempts to ramp down the current on a
significantly faster time scale are predicted to excite 2/1 tearing modes that are likely to lock to the wall, and trigger a disruption. 

The main limitation of the model presented in this paper is the use of a cylindrical plasma equilibrium with circular magnetic
flux-surfaces. The cylindrical model causes the plasma current associated with a given
value of the edge safety-factor to take  too low a value, because it does not take plasma shaping into account. The cylindrical model is also incapable of dealing with the presence of a magnetic X-point on the
plasma boundary. Other simplistic features of the model include the assumption of  uniform and constant electron number density and effective ion charge number profiles,
the neglect of the neoclassical enhancement of plasma resistivity and the bootstrap current, and the use of a prescribed, rather than a calculated, thermal diffusivity profile. 
A further deficiency of the model is that it does not include the H-L transition, which will  occur during the current ramp down in ITER \cite{muld}, or take into
account the presence of a pedestal during the H-mode phase. 
One way forward would be to combine the RAPTOR toroidal plasma equilibrium evolution code \cite{asdex,demo,muld},  the RDCON toroidal
plasma stability code \cite{rdcon}, and the SLAYER resonant layer code \cite{slayer}, in order to make realistic assessments of MHD stability during current ramps in tokamaks. 
The use of the RAPTOR code would correct all of the above-mentioned deficiencies  in the  model outlined in this paper. 

\funding{This research was funded by the  U.S.\ Department of Energy, Office of Science, Office of Fusion Energy Sciences under contract DE-FG02-04ER54742.}


\data{The digital data used in the figures in this paper can be obtained from the author upon reasonable request.}

\appendix
\section{Finite-difference solution of magnetic field equation}\label{appa}
We shall adopt a finite-difference approach, in both space and time, to calculating the poloidal magnetic field.
 Let our radial grid-points lie at $\hat{r}_i=i\,{\mit\Delta}\hat{r}$, for $i=0,I$, where ${\mit\Delta}\,\hat{r}=1/I$. Let our temporal grid-points lie at
$\hat{t}_n = n\,{\mit\Delta}\hat{t}$, for $n=0,N$.
 Discretizing Eqs.~(\ref{jdef}) and (\ref{edef}), we can write
\begin{align}
\skew{3}\hat{j}_i&= \frac{1}{\hat{r}_i\,{\mit\Delta}\hat{r}\,\delta^2}\,\left(\hat{r}_{i+1/2}\,\hat{B}_{i+1/2} -\hat{r}_{i-1/2}\,\hat{B}_{i-1/2}\right),\\[0.5ex]
\hat{E}_i &=\frac{\skew{3}\hat{j}_i}{\hat{T}_i^{\,3/2}},\\[0.5ex]
\left(\frac{\partial\hat{E}}{\partial\hat{r}}\right)_i&= \frac{1}{{\mit\Delta}\hat{r}} \left[\left(\frac{\partial\hat{E}}{\partial\hat{r}}\right)_{i+1/2}-
\left(\frac{\partial\hat{E}}{\partial\hat{r}}\right)_{i-1/2}\right]=
 \frac{a_i\,\hat{B}_{i-1}+b_i\,\hat{B}_i+c_i\,\hat{B}_{i+1}}{\delta^2\,({\mit\Delta}\hat{r})^2},\\[0.5ex]
\left(\hat{r}\,\frac{\partial\hat{B}}{\partial\hat{r}}\right)_i &= \frac{r_i}{2\,{\mit\Delta}\hat{r}}\,(\hat{B}_{i+1}-\hat{B}_{i-1}).
\end{align}
where 
\begin{align}\label{eaa}
a_i &=\frac{2}{\hat{T}_{i-1}^{\,3/2}+ \hat{T}_i^{\,3/2}}\left(1- \frac{\mit\Delta\hat{r}}{\hat{r}_{i-1}+\hat{r}_{i}}\right),\\[0.5ex]
b_i &=-\frac{2}{\hat{T}_{i+1}^{\,3/2}+ \hat{T}_i^{\,3/2}}\left(1 - \frac{\mit\Delta\hat{r}}{\hat{r}_{i+1}+\hat{r}_{i}}\right)-\frac{2}{\hat{T}_{i-1}^{\,3/2}+ \hat{T}_i^{\,3/2}}\left(1+\frac{\mit\Delta\hat{r}}{\hat{r}_{i-1}+\hat{r}_{i}}\right),\\[0.5ex]
c_i &= \frac{2}{\hat{T}_{i+1}^{\,3/2}+ \hat{T}_i^{\,3/2}}\left(1 + \frac{\mit\Delta\hat{r}}{\hat{r}_{i+1}+\hat{r}_{i}}\right).\label{ecc}
\end{align}
Here, $\hat{B}_i\equiv \hat{B}(\hat{r}_i)$,
$\hat{B}_{i+1/2} = (\hat{B}_{i+1}+\hat{B}_{i})/2$, et cetera.

The standard Crank-Nicolson algorithm\,\cite{cn} applied to Eq.~(\ref{e85}) yields
\begin{equation}
\frac{\hat{B}_i^{\,n+1} - \hat{B}_i^{\,n}}{{\mit\Delta}\hat{t}} = \frac{1}{2}\left(\frac{\partial\hat{E}}{\partial\hat{r}}\right)_i^{n+1}+\frac{1}{2}\left(\frac{\partial\hat{E}}{\partial\hat{r}}\right)_i^n+\frac{1}{2} \left[V\left(\hat{r}\,\frac{\partial\hat{B}}{\partial\hat{r}}+\hat{B}\right)\right]_i^{n+1} + \frac{1}{2} \left[V\left(\hat{r}\,\frac{\partial\hat{B}}{\partial\hat{r}}+\hat{B}\right)\right]_i^{n}.
\end{equation}
Here, $\hat{B}_i^n\equiv\hat{B}(\hat{r}_i,\hat{t}_n)$, et cetera.
Let
 $D = {\mit\Delta}\hat{t}/[2\,({\mit\Delta}\hat{r})^2]$,
$D^n = D/(\delta^n)^2$, $F^n = {\mit\Delta}\hat{t}\,V^n/(4\,{\mit\Delta}\hat{r})$, 
$\alpha_i =  -D^{n+1}\,a_i+F^{n+1}\,\hat{r}_i$,
$\beta_i = 1-D^{n+1}\,b_i - 2\,F^{n+1}\,{\mit\Delta}\hat{r}$, 
$\gamma_i =  - D^{n+1}\,c_i- F^{n+1}\,\hat{r}_i$, 
$\hat{\alpha}_i = D^n\,a_i-F^n\,\hat{r}_i$,
$\hat{\beta}_i= 1+D^n\,b_i+ 2\,F^n\,{\mit\Delta}\hat{r}$, 
$\hat{\gamma}_i = D^n\,c_i+ F^n\,\hat{r}_i$. 
 We obtain
\begin{equation}\label{trai}
\alpha_i\,B_{i+1}^{n+1} +\beta_i\,B_{i}^{n+1}+\gamma_i\,B_{i-1}^{n+1}  =\hat{\alpha}_i\,B_{i-1}^n + \hat{\beta}_i\,B_{i}^n+ \hat{\gamma}_i\,B_{i-1}^n,
\end{equation}
for $i=1, I-1$. 
Finally, the boundary conditions (\ref{bc1}) and (\ref{bc2}) give 
\begin{align}
\hat{B}_{0}^{\,n+1} &=0,\\[0.5ex]
\hat{B}_{I}^{\,n+1} &=\hat{I}_p^{\,n+1},\label{trai1}
\end{align}
where $\hat{I}_p^{\,n+1} = \hat{I}_p(\hat{t}_{n+1})$. 
Equations~(\ref{trai})--(\ref{trai1}) constitute a tridiagonal matrix system that can be solved by means of the Thomas algorithm \cite{thom}; thus, allowing the poloidal magnetic field
to be updated at each  time step. 

\section{Ramp functions}\label{apb}
Let 
\begin{equation}
F_{\rm ramp}(\hat{t},\hat{t}_0,\hat{t}_1,\hat{\tau},x) = 
1+A \left\{
 \begin{array}{lcl}
 0&&\hat{t}<\hat{t}_0\\[0.5ex]
 (\hat{t}-\hat{t}_0)^2/(2\,\hat{\tau})&&\hat{t}_0<\hat{t}<\hat{t}_0+\hat{\tau}\\[0.5ex]
 \hat{t}-\hat{t}_0-\hat{\tau}/2&&\hat{t}_0+\hat{\tau}<\hat{t} \leq \hat{t}_0+\hat{t}_1-\hat{\tau}\\[0.5ex]
 (\hat{t}_1-\hat{\tau}) - A\,(\hat{t}_0+\hat{t}_1-\hat{t})^2/(2\,\hat{\tau})&& \hat{t}_0+\hat{t}_1-\hat{\tau}< \hat{t}<\hat{t}_0+\hat{t}_1\\[0.5ex]
 \hat{t}_1-\hat{\tau}&& \hat{t}_0+\hat{t}_1<\hat{t}
 \end{array}\right.,
 \end{equation}
 where $A=(x-1)/(\hat{t}_1-\hat{\tau})$. This function ramps down, approximately linearly, from $F_{\rm ramp}=1$ to $F_{\rm ramp}=x$ between times $\hat{t}=\hat{t}_0$ and $\hat{t}=\hat{t}_0+\hat{t}_1$. 
 If $\dot{F}_{\rm ramp}\equiv\partial F_{\rm ramp}/\partial\hat{t}$ then 
 \begin{equation}
 \dot{F}_{\rm ramp}(\hat{t},\hat{t}_0,\hat{t}_1,\hat{\tau},x)= A\left\{
 \begin{array}{lcl}
 0&&\hat{t}<\hat{t}_0\\[0.5ex]
 (\hat{t}-\hat{t}_0)/\hat{\tau}&&\hat{t}_0<\hat{t}<\hat{t}_0+\hat{\tau}\\[0.5ex]
 1&&\hat{t}_0+\hat{\tau}<\hat{t} \leq \hat{t}_0+\hat{t}_1-\hat{\tau}\\[0.5ex]
 (\hat{t}_0+\hat{t}_1-\hat{t})/\hat{\tau} && \hat{t}_0+\hat{t}_1-\hat{\tau}< \hat{t}<\hat{t}_0+\hat{t}_1\\[0.5ex]
0&& \hat{t}_0+\hat{t}_1<\hat{t}
 \end{array}\right.
 \end{equation}
 Note that the finite switch-on time, $\hat{\tau}<2\,\hat{t}_1$, ensures that both $F_{\rm ramp}$ and its time derivative are continuous in time.


\begin{thebibliography}{99}\baselineskip 5ex

\bibitem{book} J.A.~Wesson, {\em Tokamaks}, 4th ed., (Oxford University Press, Oxford UK, 2011).

\bibitem{jack} G.L.~Jackson, P.A.~Politzer, D.A.~Humphreys, T.A.~Casper,  A.W.~Hyatt, J.A.~Leuer, J.~Lohr, T.C.~Luce, M.A.~Van Zeeland and J.H.~Yu, 
{\em Understanding and predicting the dynamics of tokamak discharges during startup and rampdown}, Phys.\ Plasmas {\bf 17}, 056116 (2010). 

\bibitem{imb} F.~Imbeaux, J.~Citrin, J.~Hobirk, G.M.D.~Hogeweij, F.~K\"{o}chl, V.M.~Leonov, S.~Miyamoto, Y.~Nakamura, V.~Parail, G.~Pereverzev,
et al., {\em Current ramps in tokamaks: from present experiments to ITER scenarios}, Nucl.\ Fusion {\bf 51}, 083026 (2011).

\bibitem{vries} P.C.~de Vries, T.C.~Luce, Y.S.~Bae, S.~Gerhardt, X.~Gong, Y.~Gribov, D.~Humphries, A.~Kavin,
R.R.~Khayrutdinov, C.~Kessel, et al., {\em Multi-machine analysis of termination scenarios with comparison to simulation
of controlled shutdown in ITER discharges}, Nucl.\ Fusion {\bf 58}, 026019 (2017).

\bibitem{asdex} S.~Van Mulders, O.~Sauter, C.~Contr\'{e}, F.~Felici, R.~Fischer, T.~P\"{u}tterich, B.~Sieglin,  A.A.~Teplukhina and ASDEX Upgrade Team, 
{\em Scenario optimization for the tokamak ramp-down phase in RAPTOR: Part A.\  analysis and model validation on ASDEX Upgrade}, 
Plasma Phys.\ Control.\ Fusion {\bf 66}, 025006 (2023). 

\bibitem{demo} S.~Van Mulders, O.~Sauter, C.~Contr\'{e}, E.~Fable, F.~Felici, P.~Manas, M.~Mattei, F.~Palermo, M.~Siccinio and A.A.~Teplukhina,
{\em Scenario optimization for the tokamak ramp-down phase in RAPTOR: Part B.\ safe termination of DEMO plasmas}, Plasma Phys.\ Control.\ Fusion
{\bf 66}, 025007 (2024). 

\bibitem{tcv} A.M.~Wang, A.~Pau, C.~Rea, O.~So, C.~Dawson, O.~Sauter, M.D.~Boyer, A.~Vu, C.~Galperti,  C.~Fan, et al., 
{\em Learning plasma dynamics and robust rampdown trajectories with predict-first experiments at TCV}, Nat.\ Commun.\ {\bf 16},  8877 (2025).

\bibitem{muld} S.~Van Mulders and O.~Sauter, {\em On the timescales of controlled termination of tokamak plasmas}, arXiv:2603.12972v1 (2026). 

\bibitem{fried} J.P.~Freidberg, {\em Ideal magnetohydrodynamic theory of magnetic fusion systems}, Rev.\ Mod.\ Phys.\ {\bf 54}, 801 (1982).

\bibitem{kad} B.B.~Kadomtsev, {\em Disruptive instabilities in tokamaks}, Sov.\ J.\ Plasma Phys.\ {\bf 1}, 289 (1975).

\bibitem{sweeney} R.M.~Sweeney, W.~Choi, R.J.~La Haye, S.~Mao, K.E.J.~Olofsson and F.A.~Volpe,
{\em Statistical analysis of $m/n = 2/1$ locked and quasi-stationary modes with rotating precursors at DIII-D}, 
Nucl.\  Fusion {\bf 57}, 016019 (2017). 

\bibitem{jet} G.~Sias, B.~Cannas, A.~Fanni, A.~Murari and A.~Pau, {\em Locked mode indicator for disruption prediction on JET},  
Fus.\  Eng.\  Design {\bf 138}, 254 (2019).

\bibitem{rf1} R.~Fitzpatrick,  {\em A simple model of current ramp-up and ramp-down in tokamaks},
Nucl.\ Fusion {\bf 66}, 016012 (2025).

\bibitem{cheng} C.Z.~Cheng, H.P.~Furth and A.H.~Boozer, {\em MHD stable region of the tokamak}, Plasma Phys.\ Control.\ Fusion {\bf 29}, 351 (1987).

\bibitem{hus} G.T.A.~Huysmans and O.~Sauter,
{\em MHD stability in tokamaks}, Plasma Phys.\ and Control.\ Fusion {\bf 45}, B159 (2003).

\bibitem{spitzer} L.~Spitzer, Jr., {\em Physics of fully ionized gases}, (Interscience, New York NY, 1956).

\bibitem{fitz} R.~Fitzpatrick, {\em Plasma physics: an introduction}, 2nd ed., (CRC, Boca Raton FL, 2023).

\bibitem{fitz1} R.~Fitzpatrick, {\em Tearing mode dynamics in tokamak plasmas}, (IOP, Bristol UK, 2023).

\bibitem{iter} ITER Physics Expert Group on Confinement and Transport, {\em Chapter 2: Plasma confinement and transport}, Nucl.\ Fusion {\bf 39}, 2175 (1999). 

\bibitem{ejima} S.~Ejima, R.W.~Callis, J.L.~Luxon, R.D.~Stambaugh, T.S.~Taylor and J.C.~Wesley, {\em Volt-second
analysis and consumption in Doublet III plasmas}, Nucl.\ Fusion {\bf 22}, 1313 (1982). 

\bibitem{romero} J.A.~Romero and JET-EFDA Contributors, {\em Plasma internal inductance dynamics in a tokamak}, Nucl.\ Fusion {\bf 50}, 115002 (2010). 

\bibitem{ariola} M.~Ariola and A.~Pironti, {\em Magnetic control of tokamak plasmas}, 2nd ed.\ (Springer International, Switzerland, 2016).

\bibitem{tom} G.~De Tommasi, R.~Albanese, G.~Ambrosino, M.~Ariola, P.J.~Lomas, A.~Pironti, F.~Sartori and JET-EDFA Contributors, {\em Current, position, and shape control in tokamaks}, Fus.\ Sci.\ Tech.\, {\bf 59}, 486 (2017).

\bibitem{wesson} J.A.~Wesson, {\em Hydromagnetic stability of tokamaks}, Nucl.\ Fusion {\bf 18}, 87 (1978).

\bibitem{fkr} H.P.~Furth,  J.~Killeen and M.N.~Rosenbluth,  {\em Finite‐resistivity instabilities of a sheet pinch}, Phys.\ Fluids {\bf 6}, 459 (1963).

\bibitem{ruth} P.H.~Rutherford, {\em Nonlinear growth of the tearing mode}, Phys.\ Fluids {\bf 16}, 1906 (1973).

\bibitem{ggj} A.H.~Glasser, J.M.~Greene and J.L.~Johnson, {\em Resistive instabilities in general toroidal plasma configurations}, Phys.\ Fluids {\bf 18}, 875 (1975).

\bibitem{ggj1} A.H.~Glasser, J.M.~Greene and J.L.~Johnson, {\em Resistive instabilities in a tokamak}. Phys.\ Fluids {\bf 19}, 567 (1976).

\bibitem{lut} H.~L\"{u}tjens, J.-F.~Luciani and X.~Garbet, {\em Global magnetohydrodynamic stability of tokamak plasmas with flow and resistivity}, Phys.\ Plasmas {\bf 8}, 4267 (2001).

\bibitem{rf2} R.~Fitzpatrick, {\em Investigation of tearing mode stability near ideal stability boundaries via asymptotic matching techniques}, Phys.\ Plasmas
{\bf 32}, 062509 (2025). 

\bibitem{car} R.~Carrera, R.D.~Hazeltine and M.~Kotschenreuther, {\em Island bootstrap current modification of the nonlinear dynamics of the tearing mode},
Phys.\ Fluids {\bf 29}, 899 (1986).

\bibitem{rf3} R.~Fitzpatrick, {\em Helical temperature perturbations associated with tearing modes in tokamak plasmas}, Phys.\ Plasmas {\bf 2}, 825 (1995).

\bibitem{white} R.A.~White, D.A.~Monticello, M.N.~Rosenbluth and B.V.~Waddell, {\em Saturation of the tearing mode}, Phys.\ Fluids {\bf 20}, 800 (1977).

\bibitem{has} R.J.~Hastie, F.~Militello and F.~Porcelli, {\em Nonlinear saturation of the tearing mode}, Phys.\ Rev.\ Lett.\ {\bf 95}, 065001 (2005). 

\bibitem{rf4} R.~Fitzpatrick, {\em Interaction of tearing modes with external structures in cylindrical geometry}, Nucl.\ Fusion {\bf 33}, 1049 (1993).

\bibitem{nave} M.F.F.~Nave and J.A.~Wesson, {\em Mode locking in tokamaks}, Nucl.\ Fusion {\bf 30}, 2575 (1990).

\bibitem{lh} R.J.~La Haye, C.~Paz-Soldan and Y.Q.~Liu, Nucl.\ Fusion {\bf 57}, 014004 (2017).

\bibitem{rf5} R.~Fitzpatrick, {\em Theoretical investigation of braking of tearing mode rotation by resistive walls in ITER},
Phys.\  Plasmas {\bf 30}, 042514 (2023).

\bibitem{creely} A.J.~Creely, M.J.~Greenwald, S.B.~Ballinger, D.~Brunner, J.~Canik, J.~Dooly, 
T.~F\"{u}lop, D.T.~Garnier, R.~Granetz, T.K.~Gray, et al., {\em Overview of the SPARC tokamak}, J.\ Plasma Phys.\ {\bf 86}, 865860502 (2020).

\bibitem{iter1} ITER Physics Basis Editors, {\em Chapter 8: MHD stability, operational limits and disruptions}, Nucl.\ Fusion {\bf 39}, 2137 (1999). 

\bibitem{uckam} N.A.~Uckan and the ITER Physics Group, {\em International Atomic Energy Agency, ITER Physics Design Guidelines: 1989},  (International Atomic Energy Agency, Vienna, 1990).

\bibitem{rdcon} A.H.~Glasser, Z.R.~Wang and J.-K.~Park, {\em Computation of resistive instabilities by matched asymptotic expansions},  Phys.\ Plasmas {\bf 23}, 112506 (2016).

\bibitem{slayer} J.-K.~Park, {\em Parametric dependencies of resonant layer responses across linear, two-fluid, drift-MHD regimes}. 
Phys.\ Plasmas {\bf 29}, 072506 (2022).

\bibitem{cn} J.~Crank and P.~Nicolson,
{\em A practical method for numerical evaluation of solutions of partial differential equations of the heat-conduction type}, 
Math.\ Proc.\ Cambridge Philos.\ Soc.\ {\bf 43}, 50 (1947).

\bibitem{thom} G.D.~Smith,  {\em Numerical solution of partial differential equations: finite difference methods}, 
(Oxford University Press, Oxford UK, 1965).

\end{thebibliography}
\end{document}